\documentclass[review]{elsarticle}
\usepackage{lineno,hyperref}
\modulolinenumbers[5]
\usepackage{caption}
\usepackage{subcaption}
\usepackage{graphicx}
\usepackage{graphicx}
\usepackage{xspace}
\usepackage{color}
\usepackage{braket}
\usepackage{multirow}
\usepackage{amsmath}

\definecolor{MyBlue}{rgb}{0.1,0.3,0.75}

\newcommand{\avg}[1]{\langle #1 \rangle}

\begin{document}

\begin{frontmatter}

\title{Boson Slave Solver (BoSS) v1.1 }

\author[flatiron,northwestern]{Alexandru B. Georgescu}
\author[yale]{Minjung Kim}
\author[yale]{Sohrab Ismail-Beigi\corref{corrauthor}\fnref{myfootnote}}
\cortext[corrauthor]{Corresponding author}
\ead{sohrab.ismail-beigi@yale.edu}

\address[flatiron]{Center for Computational Quantum Physics, Flatiron Institute, 162 5th Avenue, New York, New York 10010, USA}
\address[northwestern]{Department of Materials Science and Engineering, Northwestern University, 2220 Campus Drive, Evanston, Illinois 60208, United States}
\address[yale]{Department of Applied Physics, Yale University, New Haven, Connecticut 06520, USA}

\begin{abstract}

Accurate and computationally efficient modeling of systems of interacting electrons is an outstanding problem in theoretical and computational materials science.  For materials where strong electronic interactions are primarily of a localized character and act within a subspace of localized quantum states on separate atomic sites (e.g., in transition metal and rare-earth compounds), their electronic behaviors are typically described by the Hubbard model and its extensions.  In this work, we describe BoSS (Boson Slave Solver), a software implementation of the slave-boson method appropriate for describing a variety of extended Hubbard models, namely $p-d$ models that include both the interacting atomic sites (``$d$'' states) and non-interacting or ligand sites (``$p$'' states).  We provide a theoretical background, a description of the equations solved by BoSS, an overview of the algorithms used, the key input/output and control variables of the software program, and tutorial examples of its use featuring band renormalization in SrVO$_3$, Ni $3d$ multiplet structure in LaNiO$_3$, and the relation between the formation of magnetic moments and insulating behavior in SmNiO$_3$.  BoSS interfaces directly with popular electronic structure codes: it can read the output of the Wannier90 software package~\cite{mostofi_wannier90:_2008,wannier90website} which postprocesses results from workhorse electronic structure software such as Quantum Espresso~\cite{QE} or VASP~\cite{VASP}.

\end{abstract}

\begin{keyword}
electronic structure, correlated electrons, slave boson, Hubbard model, spinon
\end{keyword}

\end{frontmatter}

\textbf{Program summary}\\

Developer's repository link: bitbucket.org/yalebosscode/boss\\
Licensing provisions: Creative Commons by 4.0 (CC by 4.0)\\
Programming language: MATLAB~\cite{MATLAB2019}\\
Nature of problem: The BoSS approach, a type of slave-boson method, provides approximate solutions to interacting electron problems described by Hubbard models in a computationally efficient manner.  Hubbard models are widely used to describe materials systems with strongly localized electron-electron interactions. The interacting fermion problem is mapped onto two separate, but easier, coupled quantum problems: non-interacting fermions moving on a lattice (spinons) via tunneling between nearby atomic orbitals, and interacting slave bosons that live on individual atomic sites.  A self-consistent description of the two degrees of freedom requires matching of mean particle numbers (spinons and bosons) on each site as well as the renormalization of tunneling events for one set of particles due to the fluctuations of the other set of particles.  The method can be used to describe the interacting electronic ground state of a particular electronic configuration, or more generally it can find the minimum energy electronic configuration by searching over various symmetry broken phases (e.g., magnetic configurations, configurations with unequal occupation of nominally equivalent atomic orbitals, etc.)\\
Solution method: The spinon and slave-boson problems are each represented as Hermitian eigenvalue problems where the lowest energy (eigenvalue) state is sought.  The present implementation uses dense matrix digaonalization for the spinon problem and can use either dense or sparse matrix diagonalization for the boson problem.  Particle number matching between the two descriptions is achieved by adjustment of Lagrange multipliers which represent potential energies for the bosons:  their appropriate values are found by applying Newton's method to match spinon and boson occupancies.  Self-consistency of tunneling processes is achieved by simple fixed point iteration (solving spinon, then slave, then spinon, etc.)  Minimization of the energy uses gradient descent with adjustable step size.\\
Additional comments including Restrictions and Unusual features: Most users will prepare the input data for BoSS by running band structure calculations on a material, e.g., density functional theory (DFT) using available software packages such as Quantum Espresso~\cite{QE}.  Post processing of these calculations to create a spatially localized basis set provides the input to BoSS: most users will create the localized description by using software that transforms the electronic description into a Wannier function basis such as Wannier90~\cite{mostofi_wannier90:_2008} which BoSS interfaces with by default.  However, one can bypass this approach and  create BoSS input files manually to describe specific desired localized electron models.\\
References:\\
http://bitbucket.org/yalebosscode/boss\\
http://www.wannier.org/\\
https://www.quantum-espresso.org/\\
https://www.mathworks.com/products/matlab.html

\section{Introduction}

One of the long-standing areas of interest in condensed matter physics involves the role and effect of electron-electron interactions on the observable properties of materials.  Due to the interactions, the motion of different electrons in the material become correlated with each other in a complex manner.  Standard tools for efficient, realistic and first principles modelling of the electronic states of materials are based on single-particle (also called mean or band field) theories: one assumes that each electron moves separately in a single shared potential field, and thus each electron has a well-defined state; the shared electronic potential is created in a self-consistent manner due to the averaged inter-electronic forces created by all the electrons.  The workhorse theoretical implementation is density functional theory (DFT) \cite{hohenberg_inhomogeneous_1964,kohn_self-consistent_1965} which has had a history of success in describing many key properties of materials (stability of various crystal phases, thermodynamic and vibrational properties, a variety of chemical reactions, etc.) \cite{DFTSUCCESS}.  Extensions to DFT to deal with stronger  electronic interactions include the widely used DFT+U approach for localized interactions~\cite{Anisimov1991,anisimov_first-principles_1997}, and more generally meta-GGAs and hybrid functionals \cite{Becke1993,Perdew1996,ComparisonMeta,MetaGGALimits,ConstraintMeta,SCAN,CompareSCAN}.

Single-particle approaches do not describe the correlation of electrons explicitly in the distribution of electrons among electronic states: a single configuration consisting of independent electronic states is assumed.  However, there are electronic phenomena where the correlations lead to important effects: e.g., quasiparticle spectral weights and lifetimes, electron energy band width renormalization, and most generally excited state properties.  Materials phenomena where explicit inclusion of correlations in the calculations are important and understood to play a key role in the physics include energy band renormalization~\cite{Nekrasov2006}, unconventional superconductivity \cite{Lichtenstein2000,Mandal2017}, magnetism and colossal magnetoresistance \cite{Ramakrishnan2004,Hariki2017,Kim2015}, electronic spectroscopy of Mott insulating states~\cite{Ren2006}, metal-insulator transitions  \cite{VanadiumDioxide,Rondinelli2008}, and coupled structural and orbital symmetry breaking \cite{polarizationantoine}.  The brute force approach of simply including more electronic configurations in the calculations leads to an impractical computational cost that grows exponentially in the number of electrons.  This has led to significant research into theoretical methods that go beyond the  single-particle description in an efficient manner.  

Dynamical mean field theory (DMFT) \cite{georges_dynamical_1996,kotliar_electronic_2006} has emerged as a standard tool to describe explicitly electronic correlations in systems where localized electronic orbitals on a subset of atoms in the material dominate the electronic correlations; the method becomes {\it ab initio} when coupled to DFT (DFT+DMFT)~\cite{georges_dynamical_1996,kotliar_electronic_2006}.  This approach has been able to describe a wide range of physical phenomena that stem from localized electronic correlations~\cite{KT2004,kotliar_electronic_2006}.  To date, most DMFT calculations use adjustable parameters to describe the strength of local electronic interactions, but the parameters can now be more quantitatively justified via {\it ab initio} calculation \cite{Seth2017,cRPA}.  The application of DMFT is most obvious in cases where a material with a symmetric crystalline structure is expected to show strong electronic interaction effects: examples include insulating behavior in the high temperature paramagnetic phase where localized magnetic moments fluctuate in time (e.g., NiO above its N\'eel temperature~\cite{Ren2006}), or  where interaction-driven bandwidth and quasiparticle weight renormalization is significant, e.g., in correlated metals such as SrVO$_3$~\cite{Nekrasov2006}.  Other concepts emerging from DMFT are the site-selective transition in rare earth nickelates~\cite{Park2012}, the interrelationship between lattice and electronic degrees of freedom in transition metal oxide heterostructures~\cite{AlexPNAS}, and the physics of materials driven by the Hund's exchange interaction~\cite{JANUS}.

However, DMFT can be computationally costly when applied to systems containing multiple inequivalent correlated atomic sites, relevant to studying complex materials or heterostructures of multiple materials. The calculations can cost a significant amount of computational time and may be outside the routine budget of many research groups. For example, in the authors' experience, it takes around 1-2 minutes running on a laptop to obtain an electronic structure and a resulting band structure within DFT for the correlated metal SrVO$_3$ that has a 5 atom formula unit cell. However, a DFT+DMFT calculation of the corresponding electronic structure with sufficient accuracy to obtain a spectral function would take 500 CPU hours  within the context of a minimal model that only treats the 3 vanadium t$_{2g}$ bands explicitly in a one-shot manner without self-consistency (this order of magnitude estimate depends on the computational approach and convergence details employed). A more complex model including more bands and charge self-consistently will increase the time requirements by half an order of magnitude. Spin-orbit coupling terms that are straightforward to include in DFT can render the DFT+DMFT calculations within existing approaches close to intractable due to the well-known sign problem, although there are recent efforts to alleviate this problem~\cite{SignProblem}.  Again, these numerical estimates are for a small five-atom simulation cell.

Therefore, there have been parallel developments of methods similar to DMFT that are more approximate but much less expensive computationally. Recently, particular effort has been put into methods such as the Gutzwiller~\cite{Brinkman1970,Kotliar1986,Buenemann1997,Gutzwiller,LDAGutzwiller,Wang2010,Hugo} as well as slave-boson approaches. Since the original Kotliar-Ruckenstein slave-boson method which permitted numerical calculations at finite Coulomb interaction parameters~\cite{Kotliar1986}, a variety of new slave-boson methods have been developed and applied to real materials including the slave-rotor~\cite{Florens2002,Florens2004}, slave-spin (in multiple varieties)~\cite{DeMedici2005,Yu2012,DeMedici2014,DeMedici2017} and the rotationally invariant slave-boson~\cite{Fresard2002,Fresard1996,RotationallyInvariant} methods. Slave-boson methods of this form have  been applied to elucidate the physics of RNiO$_3$ materials~\cite{Lau2013a} with similar phenomenological predictive power to DFT+DMFT but at much lower computational cost, as well as to study Hund's physics in Fe pnictides~\cite{Yu2012,DeMedici2017}.

We proposed~\cite{georgescu_generalized_2015,georgescu_symmetry_2017} a generalized formalism based on the slave-rotor and slave-spin methods: by noticing the commonality between the two methods, we can straightforwardly build slave-boson models that allow different levels of fine-grained description of the electronic interactions (i.e., separate or aggregated description of spin and/or orbital degrees of freedom or various combinations of them).  At the same time, our formalism corrects the weak-interaction limit of the slave-rotor method. Separately, our approach allows for spontaneous symmetry breaking (e.g., ordered magnetic states).  Our approach is instantiated in the BoSS software, which this paper describes in detail.  Our paper also provides  examples of  how to use this method to reproduce physics that is normally difficult to obtain from DFT alone.

\section{General theoretical framework}

The slave boson approach used in BoSS solves, approximately, for the ground-state properties and electronic excitations of an interacting electronic system described by a Hubbard Hamiltonian.  Detailed theoretical descriptions of the approach can be found in prior publications~\cite{georgescu_generalized_2015,georgescu_symmetry_2017}, so we will briefly summarize the ideas behind the method and then focus primarily on the formalism as it connects directly to the BoSS software implementation.

A typical Hubbard Hamiltonian for interacting electrons is written a basis of localized atomic-like orbitals: each atomic site, indexed by $i$, has a set of localized orbitals indexed by $m$ and spin $\sigma\in{\pm1}$.  The Hamiltonian has the form
\begin{equation}
\hat H = \sum_{ii'mm'\sigma} t_{imi'm'\sigma} \hat c_{im\sigma}^\dag \hat c_{i'm'\sigma} + \sum_i \hat H^{(i)}_{int}\,.
\label{eq:origH}
\end{equation}
The $\hat c_{im\sigma}$ ($\hat c_{im\sigma}^\dag$) are electron annihilation (creation) field operators for the localized state $im\sigma$, the $t_{imi'm'\sigma}$ are spin-conserving tunneling (hopping) matrix elements between two localized states $im\sigma$ and $i'm'\sigma$, and the electron-electron interactions occur on each atomic site separately; the form of $\hat H^{(i)}_{int}$ will be specified further below.  (The diagonal elements $t_{imim\sigma}$, which are called the on-site energies of the localized states $im\sigma$, are included automatically in the first term of the Hamiltonian; separately, in this work, the tunneling terms do not carry an overall minus sign in front unlike other common definitions  of the Hubbard model.)   Thus, the Hubbard Hamiltonian encodes the wave-like nature of electrons via the first tunneling term in $\hat H$ (also called the hopping or kinetic term) as well as electron-electron interactions in the second term.  Solving for the ground state wave function $\ket{\Psi_0}$ of such a Hamiltonian for many electrons is very difficult and a central challenge in modern electronic structure theory: computationally efficient approximate solutions are of great interest to the research community.

\subsection{Introducing the slave bosons}
The slave-boson approach is one such approximation.  One separates the fermionic behavior from the inter-electron charged interactions by introducing a spinless charged bosonic ``slave'' degree of freedom at the atomic sites along with neutral fermion degrees of freedom with spin called spinons (i.e., one splits the original charged and spin-1/2 electron into a charged but spinless slave boson and a chargeless but spinfull fermion with spin 1/2).  The mathematical separation is given by
\begin{equation}
    \hat c_{im\sigma} = \hat f_{im\sigma} \hat O_{i\alpha} \ , \ 
    \hat c^\dag_{im\sigma} = \hat f^\dag_{im\sigma} \hat O^\dag_{i\alpha}
\end{equation}
where $\hat f_{im\sigma}$ ($\hat f^\dag_{im\sigma}$) are fermionic annihilation field operators for the spinons, and $\hat O_{i\alpha}$ ($\hat O^\dag_{i\alpha}$) lower (raise) the number of slave bosons by one.
The index $\alpha$ of the slave bosons on site $i$ describes a disjoint set of the $\{m\sigma\}$ indices belonging to that site.  Choosing how the $\{m\sigma\}$ are partitioned into the disjoint sets $\{\alpha\}$ defines the type of slave boson model being used.  For example, the coarsest model lumps all $\{m\sigma\}$ on a site into a single bosonic degree of freedom so $\alpha$ is nil and $\hat O_{i\alpha}=\hat O_i$; the most detailed model has a separate bosonic mode for each unique spin+orbital combination so $\alpha=m\sigma$.  Other models can include having two bosons per site two account for the two values of $\sigma$ while lumping all $m$ together, or alternatively having the bosons describe the $m$ states with both spin $\sigma=\pm1$ lumped together.  

The number of bosons in channel $\alpha$ ranges from zero to the maximum number of electrons $M_\alpha^{max}$ that could be accommodated by the spin+orbital combinations belonging to $\alpha$.  The matrix representation of the boson lowering operator $\hat O_{i\alpha}$ in the basis of the number of bosons is given by the $(M_\alpha^{max}+1) \times (M_\alpha^{max}+1)$ matrix
\begin{equation}
O_{i\alpha} = \left(
\begin{array}{ccccc}
0 & 1 & 0 & \ldots & 0 \\
0 & 0 & 1 & \ldots & 0 \\
\vdots & \vdots & \vdots & \ddots & \vdots\\
0 & 0 & 0 & \ldots & 1 \\
C_{i\alpha} & 0 & 0 & \ldots & 0 
\end{array}
\right)
\label{eq:Omatrix}
\end{equation}
where the choice of constants $C_{i\alpha}$ is described further below.  Further details and derivation of the structure of the operators and matrices can be found in our prior publications \cite{georgescu_generalized_2015,georgescu_symmetry_2017}.  The Hamiltonian now takes the form
\begin{equation}
\hat H = \sum_{ii'mm'\sigma} t_{imi'm'\sigma} \hat f_{im\sigma}^\dag\hat O_{i\alpha}^\dag \hat f_{i'm'\sigma}\hat O_{i'\alpha'} + \sum_i \hat H^{(i)}_{int}\,.
\label{eq:HfandO}
\end{equation}
The index $\alpha$ labels the partitioning of states $im\sigma$ while $\alpha'$ those of $i'm'\sigma$.  The main point is that the slave bosons carry the electron charge so the interaction terms $\hat H^{(i)}_{int}$ only act on the bosonic subspace.  (For the on-site contributions $im=i'm'$, we remove the $\hat O_{i\alpha}^\dag\hat O_{i\alpha}$ operator as its presence does not change anything~\cite{georgescu_generalized_2015}.)

Exact solution of the original problem posed by the Hamiltonian of Eq.~(\ref{eq:origH}) was hard enough, but the addition of new bosonic degrees of freedom on top of the fermionic spinons makes for an even harder problem.  This is because, when solving for the ground state of the Hamiltonian of Eq.~(\ref{eq:HfandO}),  one must additionally impose the constraint that the boson and fermion numbers track each other exactly at each site in order to not introduce new quantum states to the new spinon+slave problem that did not exist in the original electron-only problem: one must restrict oneself to the subspace of states $\ket{\Xi}$ in the enlarged spinon+slave Hilbert space that obey the constraint
\begin{equation}
    \hat N_{i\alpha}\ket{\Xi} = \hat n_{i\alpha}\ket{\Xi}
    \label{eq:exactconstraint}
\end{equation}
for every site $i$ and slave mode $\alpha$ because the electron charge (carried by the bosons) must follow the spin of the electron (carried by the spinons) as the particles move about the lattice.  The number operator $\hat N_{i\alpha}$ counts the number of slave bosons at site $i$ in mode $\alpha$, while the corresponding number of spinons $\hat n_{i\alpha}$ is defineed by
\begin{equation}
    \hat n_{i\alpha} \equiv \sum_{(m\sigma)\in\alpha} \hat f_{im\sigma}^\dag \hat f_{im\sigma}\,.
\end{equation}

\subsection{Approximations and self-consistent equations}
The slave boson method makes progress by separating the spinon and slave boson behaviors in order to end up with two simpler coupled problems.  Namely, the ground state $\ket{\Psi_0}$ of the Hamiltonian of Eq.~(\ref{eq:HfandO}) is approximated as a product of a spinon wave function $\ket{\psi_f}$ and a slave wave function $\ket{\phi_s}$,  $\ket{\Psi_0}\approx\ket{\psi_f}\ket{\phi_s}$.  This approximation means that we can only enforce the constraint of Eq.~(\ref{eq:exactconstraint}) on average:
\begin{equation}
    \avg{\hat N_{i\alpha}}_s = \avg{\hat n_{i\alpha}}_f
    \label{eq:avgconstraint}
\end{equation}
In addition, as explained in our prior work~\cite{georgescu_symmetry_2017}, finding the optimal spinon and slave states  corresponds to a variational minimization of the total energy functional 
\begin{multline}
    E_{tot} =  \sum_{ii'mm'\sigma} t_{imi'm'\sigma} \avg{\hat f_{im\sigma}^\dag \hat f_{i'm'\sigma}}_f \avg{\hat O_{i\alpha}^\dag \hat O_{i'\alpha'}}_s + \sum_i \avg{\hat H^{(i)}_{int}}_s\\
    - \lambda_f \left[\braket{\psi_f|\psi_f} -1\right] - \lambda_s \left[\braket{\phi_s|\phi_s} -1\right]
    -\sum_{i\alpha}h_{i\alpha}\left[\avg{\hat n_{i\alpha}}_f - \avg{\hat N_{i\alpha}}_s\right] \\
    -\sum_{im\sigma} b_{im\sigma}\left[\avg{\hat f^\dag_{im\sigma}\hat f_{im\sigma}}_f - \nu_{im\sigma}\right]
    \label{eq:Etotfull}
\end{multline}
where the shorthands for spinon and slave expectations are
\begin{equation}
    \avg{\hat X}_f \equiv \bra{\psi_f}\hat X\ket{\psi_f} \ \ , \ \ \avg{\hat Y}_s \equiv \bra{\phi_s}\hat Y\ket{\phi_s}\,.
\end{equation}
Above, four sets of Lagrange multipliers have been introduced: $\lambda_f$ and $\lambda_s$ enforce normalization of the states $\ket{\psi_f}$ and $\ket{\phi_s}$ (i.e., $\braket{\psi_f|\psi_f}=\braket{\phi_s|\phi_s}=1$), the $h_{i\alpha}$ enforce the averaged constraint of Eq.~(\ref{eq:avgconstraint}), and the ``magnetic fields'' $b_{im\sigma}$ control the spinon occupancies and ensure $\nu_{im\sigma}=\avg{\hat f^\dag_{im\sigma}\hat f_{im\sigma}}_f$.  We note that when all the constraints are obeyed, the energy $E_{tot}$ corresponds to the expectation value of the Hamiltonian $\hat H$ over the approximate product ground state $\ket{\psi_f}\ket{\phi_s}$ and is therefore a variational energy.

Minimization of $E_{tot}$ over the two wave functions $\ket{\psi_f}$ and $\ket{\phi_s}$ leads to two separate eigenvalue problems:
\begin{equation}
    \hat H_f \ket{\psi_f} = E_f\ket{\psi_f} \ \ , \ \ \hat H_s\ket{\phi_s} = E_s\ket{\phi_s}
    \label{eq:spinonandslavehpsiepsi}
\end{equation}
where the spinon Hamiltonian $\hat H_f$ is
\begin{equation}
    \hat H_f = \sum_{ii'mm'\sigma} t_{imi'm'\sigma}\avg{\hat O_{i\alpha}^\dag \hat O_{i'\alpha'}}_s \hat f_{im\sigma}^\dag \hat f_{i'm'\sigma} 
        -\sum_{i\alpha}h_{i\alpha}\hat n_{i\alpha}
    -\sum_{im\sigma} b_{im\sigma}f^\dag_{im\sigma}\hat f_{im\sigma}     
\label{eq:Hspinonlittleb}
\end{equation}
and the slave Hamiltonian $\hat H_s$ is
\begin{equation}
    \hat H_s =  \sum_{ii'mm'\sigma} t_{imi'm'\sigma} \avg{\hat f_{im\sigma}^\dag \hat f_{i'm'\sigma}}_f \hat O_{i\alpha}^\dag \hat O_{i'\alpha'} + \sum_i \hat H^{(i)}_{int}\\
    +\sum_{i\alpha}h_{i\alpha}\hat N_{i\alpha}\,. \\
\label{eq:Hslave}
\end{equation}
The two eigenvalue equations in~(\ref{eq:spinonandslavehpsiepsi}) must be solved self-consistently since averages over slave operators enter into the spinon Hamiltonian (and vice versa).  

In addition to self-consistency,  the $h_{i\alpha}$ must be adjusted to ensure that $\avg{\hat N_{i\alpha}}_s = \avg{\hat n_{i\alpha}}_f$ is obeyed.  In practice, it is very difficult to solve these equations as written because of the opposite signs with which the $h_{i\alpha}$ enter the two Hamiltonians:  increasing $h_{i\alpha}$ in $\hat H_f$ of Eq.~(\ref{eq:Hspinonlittleb}) stabilizes larger electron occupancy on site $i$ for the spinons but does the opposite for the slaves governed by $\hat H_s$ of Eq.~(\ref{eq:Hslave}).  This leads to difficulties in reaching self-consistency as well as in stabilizing broken symmetry electronic phases (e.g., magnetism)~\cite{georgescu_symmetry_2017}.

The simple solution~\cite{georgescu_symmetry_2017} is to notice that it is the sum $h+b$ that appears in $\hat H_f$ but only $h$ in $\hat H_s$: since $h$ and $b$ are independent, one can define a new variable $B=h+b$ for the spinons so that the particle matching problem is greatly simplified.  Namely, for some fixed values of $B_{im\sigma}$, one solves for the ground state $\ket{\psi_f}$ of
\begin{equation}
    \hat H_f = \sum_{ii'mm'\sigma} t_{imi'm'\sigma}\avg{\hat O_{i\alpha}^\dag \hat O_{i'\alpha'}}_s \hat f_{im\sigma}^\dag \hat f_{i'm'\sigma} 
    -\sum_{im\sigma} B_{im\sigma}f^\dag_{im\sigma}\hat f_{im\sigma} 
    \label{eq:HspinonbigB}
\end{equation}
as well as the ground state $\ket{\phi_s}$ of $\hat H_s$ of Eq.~(\ref{eq:Hslave}) self-consistently in terms of the expectations $\avg{\hat O^\dag_{i\alpha}\hat O_{i'\alpha'}}_s$ and $\avg{\hat f_{im\sigma}^\dag \hat f_{i'm'\sigma}}_f$ while the only job of the $h_{i\alpha}$ is to ensure the slave boson occupancies match the spinon occuapncies $\avg{\hat N_{i\alpha}}_s = \avg{\hat n_{i\alpha}}_f$.  One then minimizes the total energy $E_{tot}$ versus $B_{im\sigma}$ to describe the final ground state of the system.  This ``one-sided'' particle number matching is much more stable and efficient~\cite{georgescu_symmetry_2017}, and BoSS uses this ``big $B$'' approach.

A final point regards how the constants $C_{i\alpha}$ in the $\hat O_{i\alpha}$ operators of Eq.~(\ref{eq:Omatrix}) are chosen.  For an exact solution of the ground state of the interacting problem, the actual value of the $C_{i\alpha}$ is irrelevant since those entries are never accessed~\cite{georgescu_generalized_2015,georgescu_symmetry_2017}.  However, for an approximate treatment, their choice matters.  Their values are fixed by ensuring that the non-interacting limit of the spinon+slave problem matches the non-interacting limit of the original electronic problem. Namely, solving the ground state of the spinon Hamiltonian of Eq.~(\ref{eq:HspinonbigB}) should generate the same solution as solving the original Hamiltonian of Eq.~(\ref{eq:origH}) with $\hat H^{(i)}_{int}=0$.  This means that the two sets of parameters $h_{i\alpha}$ and $C_{i\alpha}$ must be adjusted when solving the non-interacting slave problem (Hamiltonian $\hat H_s$ of Eq.~(\ref{eq:Hslave}) with $\hat H^{(i)}_{int}=0$) to ensure that both $\avg{\hat O_{i\alpha}^\dag \hat O_{i'\alpha'}}_s=1$ and  $\avg{\hat N_{i\alpha}}_s=\avg{\hat n_{i\alpha}}_f$.  The resulting values of $C_{i\alpha}$ are then used without further change when solving the interacting problem.

\subsection{Specific slave-boson problem solved by BoSS}

The discussion above has described the general aspects and philosophy of the slave-boson problem underlying the BoSS software.  We now describe the specific form(s) of the Hubbard model and slave bosons used by BoSS to flesh out the method.

The type of Hubbard model solved by BoSS is a ``$pd$ model''.  The localized basis $im\sigma$ is split into two categories: (i) one subset are strongly interacting or electronically correlated ``$d$'' states with non-zero $\hat H^{(i)}_{int}\ne0$ and associated slave boson modes $\hat O_{i\alpha}$ on the correlated atomic sites $i$, and (ii) the remainder non-interacting  ``$p$'' states on uncorrelated atomic sites with no local interactions ($\hat H^{(i)}_{int}=0$) and no associated slave bosons ($\hat O_{i\alpha}=1$).  This nomenclature derives from the physics of transition metal oxide materials where the transition metals host very localized $d$ atomic orbitals for which electronic repulsions are strong, whereas the electronegative oxygen atoms that bond with and link the transition metal atoms have $2p$ orbitals that are filled with electrons and are weakly interacting. (The correlated orbitals can also refer to the localized $f$ electrons of lanthanide- or actinide-based materials.)  The $pd$ formalism used below is very much inspired by prior work using slave rotor bosons to study oxides of nickel~\cite{lau_theory_2013}.

\begin{figure}[t!]
\centering
\includegraphics[width=4.5in]{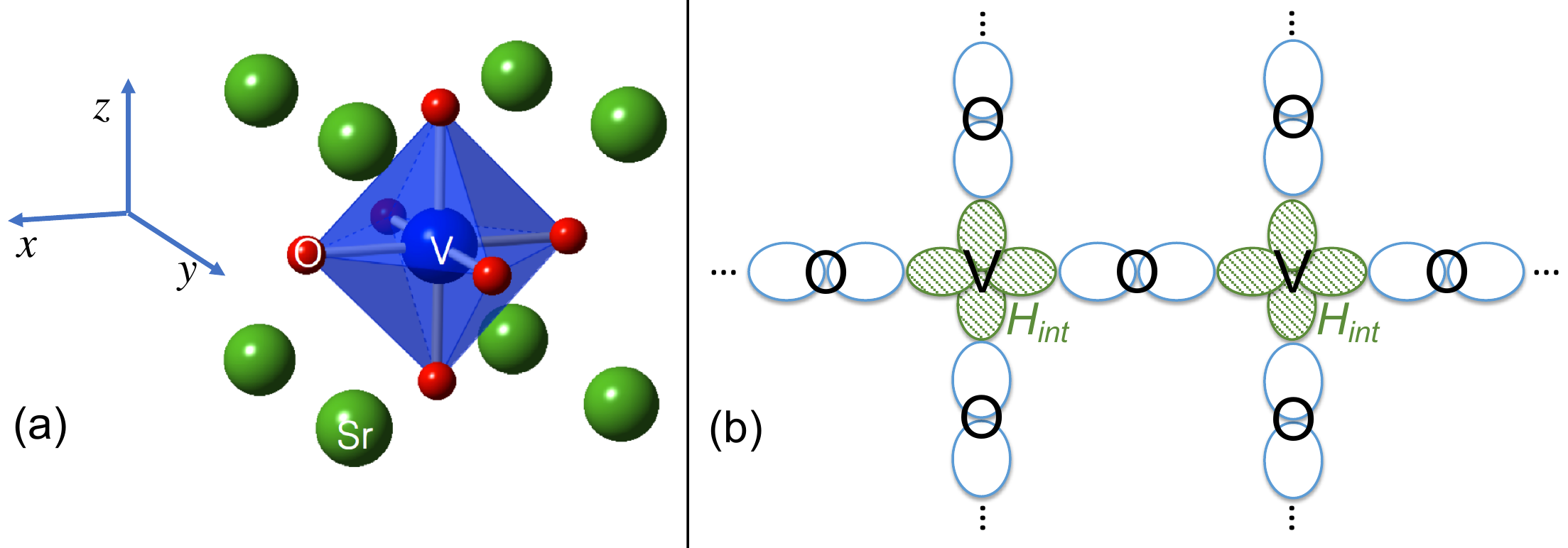}
\caption{Illustration of the typical cystal structure of transition metal oxides and the orbitals in the $pd$ model. (a) The unit cell of the cubic perovskite oxide SrVO$_3$: each V is bonded to six O atoms forming an octahedral cage (in blue); the Sr form a stabilizing cubic lattice of positive ions but do not participate significantly in the electronically conducting states.  This unit cell is periodically repeated in all three directions (i.e., a 3D tiling) to create the crystal.  (b) Schematic top view of the VO$_2$ layer in the xy plane.  Each V $d$ orbital (green hatched lobes) overlaps with its neighboring O $p$ orbitals (white lobes); the O $p$ orbitals are the bridges between neighboring V sites.  The interactions are non-zero on the correlated $d$ states, here localized on the V atoms.}
\label{fig:TMOschematic}
\end{figure}
In transition metal oxides, the transition metal atoms bond with nearest neighbor oxygen atoms.  Hence, the largest tunneling matrix elements $t_{imi'm'\sigma}$ are between the localized states of a transition metal atom and those of its oxygen neighbors.  See Figure~\ref{fig:TMOschematic} for an illustration. Thus, when constructing the slave Hamiltonian $\hat H_s$, only these nearest neighbor $t$ elements are retained.  Since the $p$ states on the oxygens do not have any associated slave modes, the slave Hamiltonian for such a $pd$ model turns into a sum of separate $d$ site Hamiltonians:
\begin{equation}
    \hat H_s = \sum_{i\in d}
     \hat H_s^{(i)}
     \label{eq:Hssumoveri}
\end{equation}
where
\begin{multline}
    \hat H_s^{(i)} = \sum_{m\sigma\in i}
    \sum_{i'm'\in p}
    \left\{[t_{imi'm'\sigma} \avg{\hat f_{im\sigma}^\dag \hat f_{i'm'\sigma}}_f] \hat O_{i\alpha}^\dag  +
    [t_{i'm'im\sigma} \avg{\hat f_{i'm'\sigma}^\dag \hat f_{im\sigma}}_f]  \hat O_{i\alpha} \right\} \\ + \hat  H^{(i)}_{int}
    +\sum_{\alpha}h_{i\alpha}\hat N_{i\alpha}\,. 
    \label{eq:Hsi}
\end{multline}
The label $d$ refers to the set of all the correlated localized states, $p$ labels all the uncorrelated localized states, $i$ is a particular correlated site with correlated states $m\sigma$, and $\alpha$ is the partitioning index of the $m\sigma$ for correlated site $i$.  We note that the spinon expectations $\avg{f_{im\sigma}^\dag \hat f_{i'm'\sigma}}_f$ renormalize the original tunneling matrix elements $t$. 

The structure of the slave-boson problem described in Eqs.~(\ref{eq:Hssumoveri},\ref{eq:Hsi})  means  that solving each correlated site $i$ separately is an exact solution to the interacting boson problem for this type of model~\cite{lau_theory_2013}.  Thus the slave ground state $\ket{\phi_s}$ is a simple product over the ground states of the separate correlated sites: $\ket{\phi_s}=\prod_{i\in d}\ket{\phi^{(i)}_s}$.

We now specify the form of the interaction part $\hat H^{(i)}_{int}$ on atomic site $i$ which can contain up to three terms depending on the specific type of slave-boson model being employed (i.e., the partitioning indexed by $\alpha$),
\begin{equation}
    \hat H^{(i)}_{int} = \hat H^{(i)}_{int,1} + \hat H^{(i)}_{int,2} + \hat H^{(i)}_{int,3}\,.
\end{equation}
Even the coarsest slave model must count the total number of slave bosons on site $i$ (i.e., a model where the $\alpha$ takes on a single value and refers to all $m\sigma$ on site $i$ so $\hat O_{i\alpha}=\hat O_i$).  Therefore, the first interaction term $\hat H^{(i)}_{int,1}$ that depends only the total boson number is always included.  It takes the form of a charging energy using a Hubbard parameter $U_i$:
\begin{equation}
\hat H^{(i)}_{int,1} =  \frac{U_i}{2} \left(\hat N_i - \avg{\hat N_i}_0\right)^2 \,.
\label{eq:Hinti1}
\end{equation}
Here, $\hat N_i = \sum_{\alpha}\hat N_{i\alpha}$ is the total number of slave modes on site $i$, and $\avg{\hat N_i}_0$ is a reference mean occupation number used for double counting corrections (see Section~\ref{sec:doublecounting}
below).  This interaction term is a charging energy that punishes charge fluctuations away from the mean value $\avg{\hat N_i}_0$.

A second interaction term $\hat H^{(i)}_{int,2}$ may be non-zero if the slave decomposition being used is able to resolve individual spatial states labeled by $m$.  In this case, one can distinguish between electronic repulsions when occupying the same orbital index $m$ with two electrons versus two different orbitals $m\ne m'$.  The added interaction term depends on an additional Hubbard parameter $U'_i$ for inter-orbital interactions:
\begin{equation}
\hat H^{(i)}_{int,2} = \frac{U'_i-U_i}{2} \left[ \left(\hat N_i - \avg{\hat N_i}_0\right)^2 - \sum_m \left(\hat N_{im} - \avg{\hat N_{im}}_0\right)^2\right]\,.
\label{eq:Hinti2}
\end{equation}
The occupation $\hat N_{im}=\sum_{\alpha|m\in\alpha} \hat N_{i\alpha}$ counts the number of bosons in spatial state $m$. An equivalent way to write this interaction term is
\begin{equation}
\hat H^{(i)}_{int,2} =\frac{U_i'-U_i}{2} \sum_{m\ne m'} \left(\hat N_{im} - \avg{\hat N_{im}}_0\right)\left(\hat N_{im'} - \avg{\hat N_{im'}}_0\right)  
\end{equation}
which shows that this interaction is a correction to the $H^{(i)}_{int,1}$ term accounting for occupation fluctuations of different spatial orbitals.

A final third term $\hat H^{(i)}_{int,3}$ is added if the  salve-boson model can resolve different spin directions $\sigma$.  This interaction represents the  classic Hund's term that lowers the energy due to same spin electron pairing on a site.  Using the Hund's interaction parameter $J$, it has the form
\begin{equation}
\hat H^{(i)}_{int,3} = -\frac{J_i}{2} \sum_\sigma  \left(\hat N_{i\sigma} - \avg{\hat N_{i\sigma}}_0\right)^2
\label{eq:Hinti3}
\end{equation}
where $\hat N_{i\sigma}=\sum_{\alpha|\sigma\in\alpha} \hat N_{i\alpha}$ counts the total number of slave boson with spin $\sigma$. 

Having specified the form of the interaction term in $\hat H_s$, the remaining matter is the choice of the $C_{i\alpha}$ in the slave $\hat O_{i\alpha}$ operators.  Since each correlated site $i$ has a separate slave Hamiltonian $\hat H^{(i)}_s$, the number of degrees of freedom are matched:  if we set $\hat H^{(i)}_{int}=0$ (i.e., $U=U'=J=0$) and solve the slave problem, we have to match two conditions $\avg{\hat O_{i\alpha}}_s=1$ and $\avg{\hat N_{i\alpha}}_s=\avg{\hat n_{i\alpha}}$ with two free parameters $h_{i\alpha}$ and $C_{i\alpha}$.  This concludes the theoretical specification of the BoSS slave problem.

The spinon Hamiltonian for the BoSS $pd$ model takes the form
\begin{multline}
    \hat H_f = \sum_{im\sigma\in d}\sum_{i'm'\in p} 
    \left\{[t_{imi'm'\sigma}\avg{\hat O_{i\alpha}^\dag}_s] \hat f_{im\sigma}^\dag \hat f_{i'm'\sigma}   +
    [t_{i'm'im\sigma} \avg{\hat O_{i\alpha}}_s ] \hat f_{i'm'\sigma}^\dag \hat f_{im\sigma} \right\}  \\
    + \left[\sum_{im\sigma\in d}\sum_{i'm'\in d} + \sum_{im\sigma\in p}\sum_{i'm'\in p} \right]
    \left\{t_{imi'm'\sigma} \hat f_{im\sigma}^\dag \hat f_{i'm'\sigma}\right\}
   \\
    -\sum_{im\sigma} B_{im\sigma}f^\dag_{im\sigma}\hat f_{im\sigma}\,. 
    \label{eq:HfBoSS}
\end{multline}
As explained above, the only modifications to the original tunneling elements $t$ are those between $p$ and $d$ localized states (the  factors of $\avg{\hat O_{i\alpha}}_s$ above).  The remainder of the tunneling matrix elements are unchanged.   This concludes the theoretical specification of the  BoSS spinon problem.

\subsection{Periodic systems and Bloch states}

The above formalism is applicable to both isolated systems such as molecules as well as extended materials such as crystalline solid state materials.  However, for crystalline systems which have a periodic arrangement of atoms over macroscopic length scales, one typically describes them using periodic boundary conditions which then permits use of Bloch's theorem to greatly reduce the size of the problem: one can replace a large simulation cell with periodic boundary conditions by instead dealing with the much smaller  primitive unit cell under ``twisted'' boundary conditions.  In the solid state language, one uses $k$-sampling over a grid of uniform grid of Bloch wave vectors $k$ in the first Brillouin zone (Born-von Karman boundary conditions)~\cite{ashcroft_solid_1976}.

Within the BoSS approach, only the spinons
are aware of the $k$-sampling because the slave problem is solved in a completely localized manner, i.e., one site at a time, and is thus unaffected by the long-range electronic boundary conditions.  For the spinons, each $k$ vector is associated with its own Hamiltonian
\begin{multline}
\hat H_f^{(k)} =  \sum_{im\sigma\in d}\sum_{i'm'\in p} 
    \left\{[t^{(k)}_{imi'm'\sigma}\avg{\hat O_{i\alpha}^\dag}_s] \hat f_{im\sigma}^\dag \hat f_{i'm'\sigma}   +
    [t^{(k)}_{i'm'im\sigma} \avg{\hat O_{i\alpha}}_s ] \hat f_{i'm'\sigma}^\dag \hat f_{im\sigma} \right\}  \\
    + \left[\sum_{im\sigma\in d}\sum_{i'm'\in d} + \sum_{im\sigma\in p}\sum_{i'm'\in p} \right]
    \left\{t^{(k)}_{imi'm'\sigma} \hat f_{im\sigma}^\dag \hat f_{i'm'\sigma}\right\}
   \\
    -\sum_{im\sigma} B_{im\sigma}f^\dag_{im\sigma}\hat f_{im\sigma}\,,
\label{eq:Hfk}
\end{multline}
where the sums over $im\sigma$ and $i'm'\sigma$ now run only over the localized states in a single unit cell, and 
\begin{equation}
t^{(k)}_{imi'm'\sigma} = \sum_R e^{ik\cdot R}\, t_{imi'm'\sigma}\,,
\label{eq:tk}
\end{equation}
and $R$ sums over the lattice vectors identifying all the primitive unit cells inside the periodic supercell.  Spinon averaged quantities are also averaged over the $k$ points:  e.g., the average $\avg{\hat X}_f$ is given by $N_k^{-1}\sum_k \avg{\hat X}_f^{(k)}$, where $\avg{\hat X}_f^{(k)}$ is the average over the ground state of $
\hat H_f^{(k)}$ and $N_k$ is the number of $k$ points.

\subsection{Relation to prior work, double counting correction}
\label{sec:doublecounting}

The formalism above differs from our prior work~\cite{georgescu_generalized_2015,georgescu_symmetry_2017} in two ways.  The minor difference is that the above BoSS approach aims to solve for the ground state of a $pd$ Hubbard model, while the prior work states the problem generally or applies it to a simpler $d$ only model where all the  localized states are correlated and centered on transition metal sites.  This boils down primarily to differences in notation and the factors involved in the rescaling of the tunneling terms in $\hat H_f$ and $\hat H_s$.  The major difference is that (a) all the electron-electron interaction terms in BoSS are contained only in the slave boson sector of the problem, and (b) there are references occupation values such as $\avg{\hat N_i}_0$ in the interaction terms.  

The two major differences stem from how the BoSS approach should be used in practice.  Our BoSS approach is aimed to be used as a post processing step to a mean field band structure calculation based on, e.g., DFT.  Namely, the BoSS model takes as input the DFT description and then tries to correct its deficiencies.  It assumes that electron-electron interactions at the mean field level, where electrons interact via averaged potentials and thus the description is of the single-particle type, are already included in the $t_{imi'm'\sigma}$ values.  Hence, the interactions that are missing from the mean-field approach are those due to fluctuations in the number of electrons on the correlated sites as described by the slave-boson part.  However, since the mean field approach already describes certain types of electron-electron interactions, we want to avoid including these interactions twice and erroneously double counting them.  

Double counting corrections have a long history and are an important part of any approach using localized basis sets for interacting electron problems~\cite{kotliar_electronic_2006}.  In the end, one posits a physically motivated correction that is exact in some limit.  For BoSS, the interactions terms in Eqs.~(\ref{eq:Hinti1},\ref{eq:Hinti2},\ref{eq:Hinti3}) are written an explicit form showing that they are non-zero when the electron number in set of localized correlated states fluctuates away from an average value such as $\avg{\hat N_i}_0$.  Physically, we expect that the slave boson theory should give no corrections to the mean field description when the electron number fluctuations about the mean field values are zero.  Hence, we choose the double counting reference electron occupations  $\avg{\hat N_i}_0$, $\avg{\hat N_{im}}_0$ and $\avg{\hat N_{i\sigma}}_0$ in Eqs.~(\ref{eq:Hinti1},\ref{eq:Hinti2},\ref{eq:Hinti3}) to be those obtained from solving the BoSS problem with no added interactions, i.e., with $U_i=U'_i=J_i=0$ or $\hat H^{(i)}_{int}=0$.

While BoSS has been designed to be a post processor for a mean field calculation in order to add missing Hubbard-type physics, one can easily use the BoSS framework to (approximately) solve a Hubbard model itself.  One simply sets the $\avg{\hat N_i}_0=0$ in the interaction terms of Eqs.~(\ref{eq:Hinti1},\ref{eq:Hinti2},\ref{eq:Hinti3}) and proceeds to solve the resulting problem.

\section{Algorithms used in BoSS}

Before describing the software implementation of BoSS, it is helpful to describe briefly the numerical algorithms used by BoSS to solve the slave-boson problem.  Describing the algorithms first helps set the stage for the the necessarily more detailed and low-level software implementation description.

The most basic problem BoSS must solve over and over is the computation of the ground state expectations of the spinon density matrix $\avg{\hat f^\dag_{im\sigma}\hat f_{i'm'\sigma}}_f$ and the slave expectation $\avg{\hat O^\dag _{i\alpha}\hat O_{i'\alpha'}}_s$.  While formally these expectations are for the ground state wave function of $\hat H_f$ (Eq.~\ref{eq:Hfk}) and $\hat H_s$ (Eq.~\ref{eq:Hsi}), respectively, in practice we use a low but finite temperature Boltzmann distribution to compute them: thermal averaging naturally averages over degenerate manifolds, provides numerical stability for near degenerate states, and accelerates sampling of the Fermi surface for metallic spinon systems.  For the non-interacting spinon Hamiltonian of Eq.~(\ref{eq:Hfk}) at a $k$ point, BoSS sets up a square hermitian Hamiltonian matrix $H^{(k,\sigma)}$ for each spin channel $\sigma$ with off diagonal $(d,p)$ entries given by $t^{(k)}_{imi'm'\sigma}\avg{O_{i\alpha}}_s$,  other off digonal entries $t^{(k)}_{imi'm'\sigma}$, and diagonal entries $t^{(k)}_{imim\sigma}-B_{im\sigma}$.  BoSS diagonalizes this matrix to obtain the band energies $\epsilon^{(k,\sigma)}_{n}$ and orthonormal eigenvectors $u_{im,n}^{(k,\sigma)}$.  The expectation is then computed using the Fermi-Dirac distribution via
\begin{equation}
    \avg{f^\dag_{im\sigma}f_{i'm'\sigma}}_f = \frac{1}{N_k}\sum_{k} \frac{u_{im,n}^{(k,\sigma)}\ {u_{i'm',n}^{(k,\sigma)}}^*}{1+\exp[-\beta(\epsilon_{n}^{(k,\sigma)}-\mu)]}\,,
\end{equation}
where $\beta=1/(k_BT)$ is the inverse thermal energy.  The chemical potential $\mu$ is determined by ensuring the correct mean number of total electrons $N_e$ per simulation cell,
\begin{equation}
    N_e = \frac{1}{N_k}\sum_{k,\sigma} \frac{1}{1+\exp[-\beta(\epsilon_{n}^{(k,\sigma)}-\mu)]}\,.
\end{equation}
The unique value of $\mu$ is determined efficiently by the  bisection algorithm~\cite{press_numerical_2007} since the summand is monotonically increasing in $\mu$.

For the slave Hamiltonian operator on each site, $\hat H_s^{(i)}$ of Eq.~(\ref{eq:Hsi}), the corresponding Hamiltonian matrix is computed in the number representation where the $\hat O_{i\alpha}$ operators have the matrix elements given by Eq.~(\ref{eq:Omatrix}), and the slave number operators $\hat N_{i\alpha}$ are diagonal matrices.  Diagonalization of the Hamiltonian produces eigenenergies $E^{(i)}_n$ and eigenstates $\ket{n^{(i)}}$ that are used to compute averages of any slave-based operator $\hat X^{(i)}$  on site $i$ via
\begin{equation}
    \avg{\hat X^{(i)}}_s = \frac{1}{Z^{(i)}}\sum_n \bra{n^{(i)}}\hat X^{(i)}\ket{n^{(i)}}e^{-\beta E_n^{(i)}} \ \ , \ \ Z^{(i)} = \sum_n e^{-\beta E_n^{(i)}}\,.
\end{equation}
Given the sparsity of Eq.~(\ref{eq:Omatrix}), the slave Hamiltonian is also sparse so BoSS can employ sparse matrix methods to store and diagonalize the Hamiltonian thereby saving signficant memory and computational effort.  In addition, since only states with a few $\beta^{-1}$ of the lowest energy contribute to the thermal averaging, the diagonalization needs only return a small subset of the lowest energy eigenvalues and associated eigenvectors.

The next higher level problem BoSS must attack is finding the lowest energy state of the slave Hamiltonian $\hat H_s^{(i)}$ of  Eq.~(\ref{eq:Hsi}) while matching certain conditions which always include matching specified spinon occupancies $\avg{\hat n_{i\alpha}}_f$.  The first case is that one is seeking to find the constants $C_{i\alpha}$ that are needed to define the $O_{i\alpha}$ matrices: one adjusts both $C_{i\alpha}$ and $h_{i\alpha}$ to match $\avg{\hat N_{i\alpha}}_s=\avg{\hat n_{i\alpha}}_f$ as well as ensure that $\avg{\hat O_{i\alpha}}_s=1$.  Due to the lack of interactions, each channel $\alpha$ can be solved separately so this represents a two-dimensional search in $(h_{i\alpha},C_{i\alpha})$ to match two conditions.  The second case is that one is solving the interacting slave-boson problem in which case the different $\alpha$ bosons on the same site $i$ are coupled so that one must search over the entire set of $\{h_{i\alpha}\}$ at each site $i$ to match all the spinon occupancies $\avg{\hat n_{i\alpha}}_f$.  Our experience shows that due to the relatively well behaved nature of both cases, a modified Newton's algorithm is sufficient to efficiently solve both problems.  Both problems are of the generic form $f(x)-y=0$ where we search for a vector $x$ that satisfies the equation for a fixed vector $y$.  The derivative matrix $df|_x$ is computed numerically by finite differences, and our modified Newton algorithm for going from Newton step $j$ to $j+1$ is
\begin{equation}
x_{j+1}=x_j - \alpha_{j+1} (df|_x)^{-1} (f(x_j)-y)\,.
\end{equation}
The scaling factor $\alpha_j=1$ defines the textbook Newton's algorithm.  However, to avoid instability and overshooting, we dynamically update $\alpha_j$ based on progress toward a solution which is based on the size of the residual $e(x)\equiv \|f(x)-y\|$ (the standard Euclidean norm).  If $e(x)$ is worsened compared to the previous step, i.e., $e(x_j)> e(x_{j-1})$, then  we reduce $\alpha_{j+1} = \alpha_j / 3$ to take a conservative small step towards the solution.  But if $e(x_j)> e(x_{j-1})$, we instead push $\alpha_{j+1}$ towards unity via $\alpha_{j+1}=2\alpha_j(2-\alpha_j)$. The computationally costly part of this approach is evaluation of the derivative matrix $df|_x$ when many boson modes $\alpha$ exist on a site.  To gain efficiency, BoSS will calculate the $df|_x$ matrix once, use it for some user-specified number of Newton steps before recomputing it (i.e., Picard's method instead of Newton's method for the intermediate steps).

One level higher is to solve the spinon+slave problem self-consistently for some  specified set of ``big $B$'' values $B_{im\sigma}$.  We have found this numerical problem to be  {\it suprisingly smooth}: a simple fixed point iteration algorithm is sufficient for rapid convergence.  Namely, given some state of the spinon+slave system at fixed $B_{im\sigma}$, BoSS uses the current spinon averages $\avg{\hat f_{im\sigma}^\dag\hat f_{i'm'\sigma}}_f$ to set up and solve the slave problem over all correlated sites $i$ which provides updated averages $\avg{\hat O_{i\alpha}}_s$; then, these updated averages are used to set up and solve the spinon problem and to update the $\avg{\hat f_{im\sigma}^\dag\hat f_{i'm'\sigma}}_f$; and the process is repeated until the magnitude the successive changes of the spinon occupancies $\avg{\hat f_{im\sigma}^\dag\hat f_{im\sigma}}_f$ over the correlated sites drop below a tolerance value.

At the highest level, BoSS must minimize the total energy $E_{tot}$ of Eq.~(\ref{eq:Etotfull}) over the $B_{im\sigma}$.  When all required constraints are met and the BoSS $pd$ approach is used, $E_{tot}$ takes the simpler form
\begin{equation}
   E_{tot}(B) = 
        \sum_{ii'mm'\sigma} t_{imi'm'\sigma} \avg{\hat f_{im\sigma}^\dag \hat f_{i'm'\sigma}}_f \avg{\hat O_{i\alpha}^\dag}_s\avg{\hat O_{i'\alpha'}}_s + \sum_i \avg{\hat H^{(i)}_{int}}_s
      \label{eq:Etotused}
\end{equation}
where vector $B$ contains all the $B_{im\sigma}$ values. The $B$-dependence of $E_{tot}$ comes from the spinons via their Hamiltonian of Eq.~(\ref{eq:HfBoSS}).  BoSS minimizes this energy by simple gradient descent in $B$ with adjustable step size. The gradient $\nabla_B E_{tot}(B)$ is computed numerically by finite differences of the components of $B$.  The update step is $B_{j+1}=B_j - \gamma_{j+1}\nabla_B E_{tot}(B_j)$.  The scaling factor $\gamma_j>0$ is adjusted based on progress in lowering the energy.  If the energy went down, i.e.,  $E_{tot}(B_{j})<E_{tot}(B_{j-1})$, then the step size is increased via $\gamma_{j+1}=G\gamma_j$ with a growth factor $G>1$ (the default value is $G=1.1$).  However, if the energy went up compared to the previous step, the step size is reduced via $\gamma_{j+1}=\gamma_j / R$ with $R>1$ (the default value is $R=3$).  This simple algorithm attempts to adjust the steps in $B$ to be as large as possible while still decreasing $E_{tot}$.  The minimization is terminated when successive changes of $E_{tot}$ are below tolerance.

\section{Software implementation}

The BoSS software has been implemented is in the MATLAB~\cite{MATLAB2019} programming and software environment.  This environment is widely available on many computational platforms and allows for rapid software development, testing, as well as plotting and visualization.  In what follows, we briefly describe the program flow, input/output and key variables in BoSS.  File names or key variables names associated with a particular routine or setting are typeset as \verb+filename+ or \verb+variablename+ below.  

\subsection{BoSS program flow}

\begin{figure}
\centering
\includegraphics[width=5in]{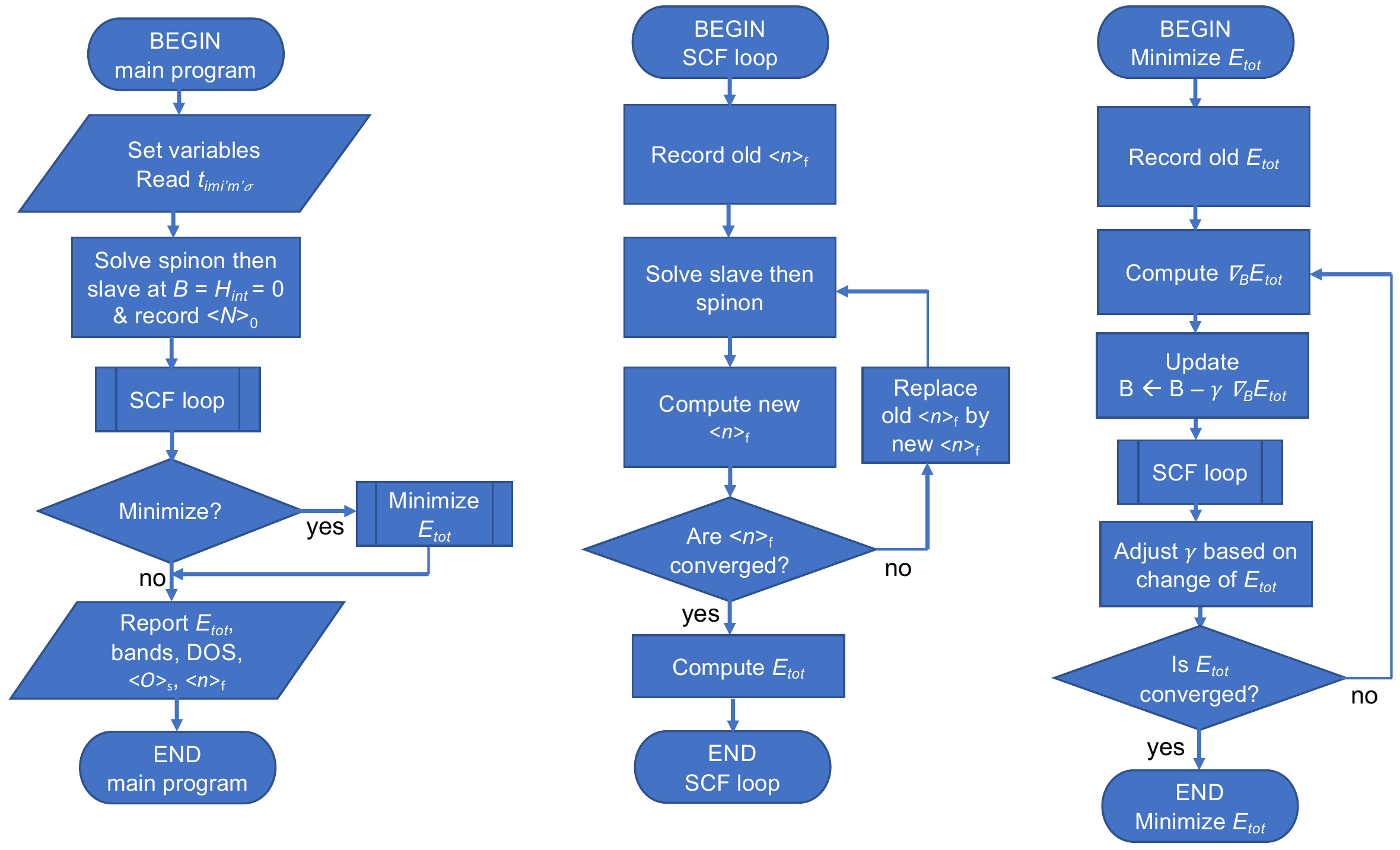}
\caption{High level schematic flowchart of the main control flow in BoSS.  The main program flow (left column) relies on two subroutines: ``SCF loop'' (defined in the center column) and ``Minimize $E_{tot}$'' (defined in the right column which also relies on SCF loop).}
\label{fig:flowchart}
\end{figure}

The main program file \verb+mainprogram.m+ and important subroutine file \verb+setup_system.m+ (that reads the input data) reside in the top level directory of the BoSS package while the remaining subroutine are in a  \verb+functions/+ subdirectory.  A high level overview of the software is provided by the flowchart in Figure~\ref{fig:flowchart}.  

The main program (\verb+mainprogram.m+) calls a subroutine to initialize key variables (\verb+setup_system.m+) and then does a self-consistent field (SCF) calculation of the slave-boson problem (\verb+functions/SCFloop.m+) before reporting on the solution; if requested, the main flow calls the  \verb+functions/minimize_Etot.m+ subroutine before reporting on final results.  The minimization of $E_{tot}$ via gradient descent also relies on \verb+functions/SCFloop.m+ to find a self-consistent solution at a given value of the $B_{im\sigma}$ variables.  The SCF loop implements a simple self-consistency loop over the spinon occupations of the correlated orbitals by calling lower level routines which use the algorithms discussed in the previous section.  

The main computational subroutines that these high level activities depend on are:
\begin{itemize}
\item \verb+functions/solve_spinon_then_slave.m+ and\\
 \verb+functions/solve_slave_then_spinon.m+ : these two similar routines solve the spinon and slave problems in the order specified by their file names.
\item \verb+functions/slave_driver.m+ : loops over correlated sites and solves the slave problem on each site while matching spinon occupancies.
\item \verb+functions/Csearch.m+ : solves the slave problem on a site at zero interaction strength to find the $C_{i\alpha}$ values for that site that give $\avg{O_{i\alpha}}_s=1$ (while also matching spinon occupancies using the $h_{i\alpha}$).
\item \verb+functions/hsearch.m+ : solves the interacting slave problem on a site while matching spinon occupancies by adjusting the $h_{i\alpha}$.
\item \verb+functions/HamSlave.m+ : low level computational routine that sets up the interacting slave-boson Hamiltonian problem on a given site and finds the ground state via diagonalization.
\item \verb+functions/solvespinon_fixedN+ : solves the spinon problem over all $k$ vectors for a fixed number of electrons.
\item \verb+functions/buildHspinon+ , \\
 \verb+functions/diagHspinon+ , and\\
 \verb+functions/diagH_kspinon+ : A set of routines that loop over the $k$ vectors, build the spinon Hamiltonian at each $k$, and then diagonalize them to find the spinon eigenvalues and eigenvectors at each $k$.
\item \verb+functions/calcrhospinon.m+ and\\
\verb+functions/findmu.m+ : compute the spinon density matrix $\avg{\hat f^\dag_{im\sigma}\hat f_{i'm'\sigma}}_f$ by summing over the $k$ and using the eigenvectors at each $k$ together with the Fermi-Dirac occupancies computed using the spinon eigenvalues and chemical potential $\mu$.
\end{itemize}

\subsection{Input hopping/tunneling elements}

The most important input to BoSS is the localized orbital (tight-binding) model specified by the tunneling matrix elements $t_{imi'm'\sigma}$. These are read from a plain text file in the format output by the Wannier90 software package~\cite{mostofi_wannier90:_2008,wannier90website} for computing maximally localized Wannier functions~\cite{marzari_maximally_1997,souza_maximally_2001,marzari_maximally_2012}.  The tight-binding data file output by Wannier90 is a plain text file named \verb+<base>_hr.dat+ (where \verb+<base>+ is a placeholder for a name chosen by the Wannier90 user).  Typically, the Wannier90 program is run as a post-processing step to a first principles DFT calculation to produce a localized basis describing the electronic structure.  However, one can generate a hand-written \verb+<base>_hr.dat+ file to describe some desired tight-binding problem (see the explanation of the tutorials in Sec.~\ref{sec:tutorial} below).

The text file \verb+<base>_hr.dat+ is generally quite long and therefore slow to process, so BoSS requires that the user perform a one time preprocessing of this file to convert it into MATLAB binary form for rapid read access.  This is accomplished by the supplied \verb+convert_hrdat_to_bin.m+ function that can process v1.1 or v1.2 formatted Wannier90 \verb+<base>_hr.dat+ files to produce the binary version.  It is the binary files that are read by the subroutine \verb+setup_system.m+ during the execution of the BoSS program.  In fact, BoSS reads two files of tight-binding data since there are two independent spin channels ($\sigma=\pm1$ or ``up''/``down'' spin): the file names are set by the variables \verb+hrbinfileup+ and \verb+hrbinfiledn+ in \verb+setup_system.m+.  This allows one to deals with spin-polarized tight-binding representations;  if no spin polarization is evident (or desired), one simply makes the two file names identical.

\subsection{Key input/control variables}

BoSS has a large number of input and control variables that are defined and set to various values in \verb+setup_system.m+.  We refer the reader to examples in the software package for a full, commented list of the variables.  Here, we highlight the meaning and implications of the more important variables.  The BoSS programming philosophy is that all input or control variables are defined and initialized in the file \verb+setup_system.m+: the rest of the program, subroutines, and functions should not contain other such variables or arbitrary numerical values (which have significant influence over the program execution or output).

The important high-level variables are common to both spinon and slave problems are:
\begin{itemize}
\item \verb+corbs+ and \verb+porbs+ : two integer arrays containing lists of localized orbitals that are correlated and uncorrelated (i.e., interacting and non-interacting), respectively.  The numbering of orbitals is that of the input Wannier90 representation.
\item \verb+occtol+ : main electron occupancy tolerance for self-consistency and number matching.  This value is used to decide if the SCF loop is converged (when \verb+corb+ spinon occupancies change by less than this magnitude between successive iterations) as well as the maximum difference allowed between slave and spinon occupancies when searching over $h_{i\alpha}$.
\item \verb+tijtol+ : tunneling elements $t_{imi'm'\sigma}$ smaller in magnitude that this number (in eV) are set to zero.  This is useful for reducing significantly the size of the tight-binding representation which typically contains many small entries between spatially far apart orbitals.  However, it may change the non-interacting spinon bands away from the ones defined by the Wannier90 output.  Setting this to zero retains all input tunneling elements.
\item \verb+minimize_Etot_over_Bfield+ : a flag deciding if minimization of $E_{tot}$ over  $B_{im\sigma}$ is to be performed (a non-zero value turns it on).
\end{itemize}

Most of the variables controlling the spinon behavior are members of the structure \verb+spinoninfo+. The key ones are:
\begin{itemize}
    \item \verb+spinoninfo.dim+ : controls the dimensionality of the $k$-sampling.  If equal to 2, $k$ vectors sample only the $xy$ plane; if equal to 3, $k$ vectors sample in all three spatial directions.
    \item \verb+spinoninfo.nk+ : the number of evenly-spacked $k$ samples along each axial direction being sampled.  The sampling directions are along the primitive reciprocal lattice vectors.
    \item \verb+spinoninfo.kT+ : temperature (in eV) for the Fermi-Dirac distribution converting spinon energies to occupancies.
    \item \verb+spinoninfo.Ne+ : the total number of spinons (i.e., electrons) in each unit cell.  This is the value the chemical potential $\mu$ search targets.
    \item \verb+spinoninfo.Bfield+ : initial values of the $B_{im\sigma}$ (in eV) that control the spinon occupancies.  These are updated if minimization is turned on.
\end{itemize}
The key variables controlling the slave bosons are members of \verb+slaveinfo+:
\begin{itemize}
    \item \verb+slaveinfo.nsites+ : the number of correlated sites.
    \item \verb+slaveinfo.nslavespersite+ : the number of slave modes per site
    \item \verb+slaveinfo.allowedOccs+ : an integer array specifying the set of allowed slave occupancy numbers on a correlated site.  For example, if a single boson describes the occupancy of entire $d$ shell, which has 5 spatial orbitals and two spin channels, then  set this to \verb+[0:10]+; in the other extreme of each boson describing a unique spin+orbital combination, set this to \verb+[0:1]+.
    \item \verb+slaveinfo.ncorbsperslave+ : the number of spatial orbitals per slave mode.  If the value is one, then the slave model can resolve individual spatial orbitals and the value of $U'$ 
    is used in the interaction Hamiltonian.
    \item \verb+slaveinfo.spinresolved+ : if set to one, the slave modes can distinguish the two spin indices $\sigma$, and this turns on the use of $J$ and the Hund's  interaction term (setting to zero turns this off).
    \item \verb+spinoninfo.U+ , \verb+spinoninfo.Up+ , \verb+spinoninfo.J+ : arrays specifying the $U,U',J$ values (in eV) for each correlated site.
    \item \verb+slaveinfo.Oavgtol+ : the tolerance within which $\avg{O_{i\alpha}}=1$ when solving the non-interacting slave problem for the $C_{i\alpha}$.
    \item \verb+slaveinfo.kTslave+ : temperature (in eV) for the Boltzmann distribution used to compute the slave-boson averages.
\end{itemize}
When minimization is performed, the structure \verb+miniminfo+ contains the variables controlling the minimization.  The most critical variable is the energy tolerance \verb+miniminfo.Etottol+ (in eV) for changes of $E_{tot}$ during minimization: when the successive change of $E_{tot}$ between gradient descent steps drops below this tolerance, the minimization is terminated.

\subsection{Key working variables}

The BoSS program flow has a number of variables that are modified as the final self-consistent and/or minimized solution is computed.  Here we focus on four basic and key variables, and reader may consult the software package for other variables and how they are computed or used.  The variables of interest are:

\begin{itemize}
    \item \verb+dcount+ : a $2\times n_c$ array containing the spinon occupancies $\avg{\hat f^\dag_{im\sigma}\hat f_{im\sigma}}_f$ of the correlated localized orbitals where $n_c$  is the length of the \verb+corb+ array (i.e., the number of spatial  orbitals that are localized).  The rows refer to the spin index $\sigma$ and the columns to the spatial orbitals in the order specified in \verb+corb+.
    \item \verb+Oavg+ : a $2\times n_c$ array containing the slave averages $\avg{\hat O_{i\alpha}}_s$.  Correlated localized states belong to the same $i\alpha$ index have the same \verb+Oavg+ values.
    \item \verb+Eint+ : expectation value of the total electron-electron interaction energy, the second term on the right hand side of Eq.~(\ref{eq:Etotused}).
    \item \verb+Eband+ : expectation value of the hopping energy, the first term on the right hand side of Eq.~(\ref{eq:Etotused}).
    \item \verb+Etot+: the sum \verb|Eband + Eint|.
\end{itemize}

\section{Tutorial examples}
\label{sec:tutorial}

The BoSS software package is distributed with four examples forming an introductory tutorial.  The first example is about the electronic structure of SrVO$_3$, a metallic and non-magnetic cubic perovskite transition metal oxide whose observed electronic bands show significant quantitative differences from the DFT-calculated ones for a 5-atom primitive unit cell.  The second example is about how one can  create a Wannier90-formatted \verb+<base>_hr.dat+ file easily to describe a desired Hubbard model.  The third example shows the effect of having the $B_{im\sigma}$ symmetry breaking fields, and how they can be determined via minimization of the total energy $E_{tot}$. The fourth examples shows how comparing two different slave models for the same material, LaNiO$_3$, can give insight into the key physics.  We will summarize key aspects of the examples below, and refer the reader to the software package's tutorial documentation and downloadable files for full details.
\begin{figure}[t!]
\begin{centering}
\includegraphics[width=2.3in]{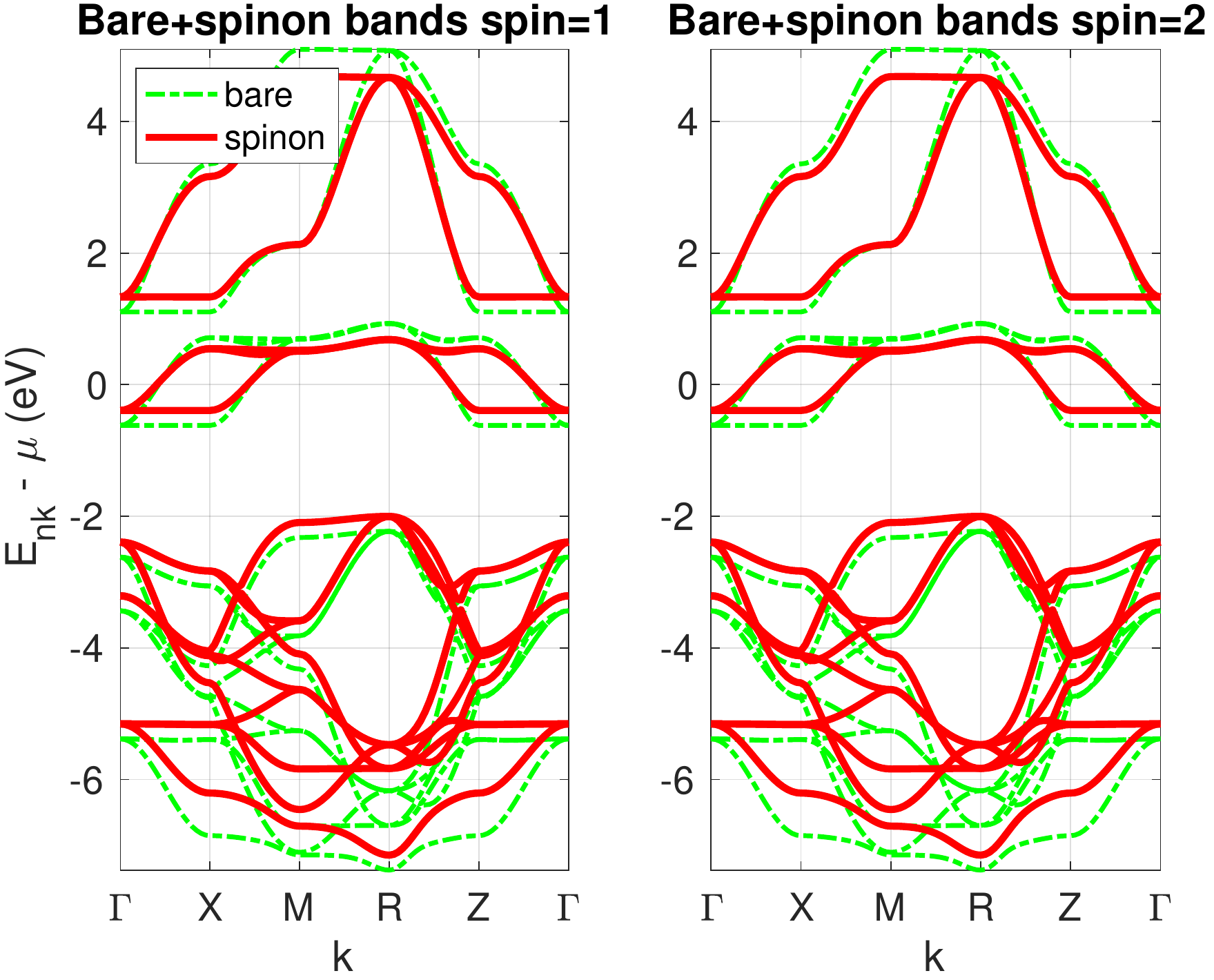}
\includegraphics[width=2.3in]{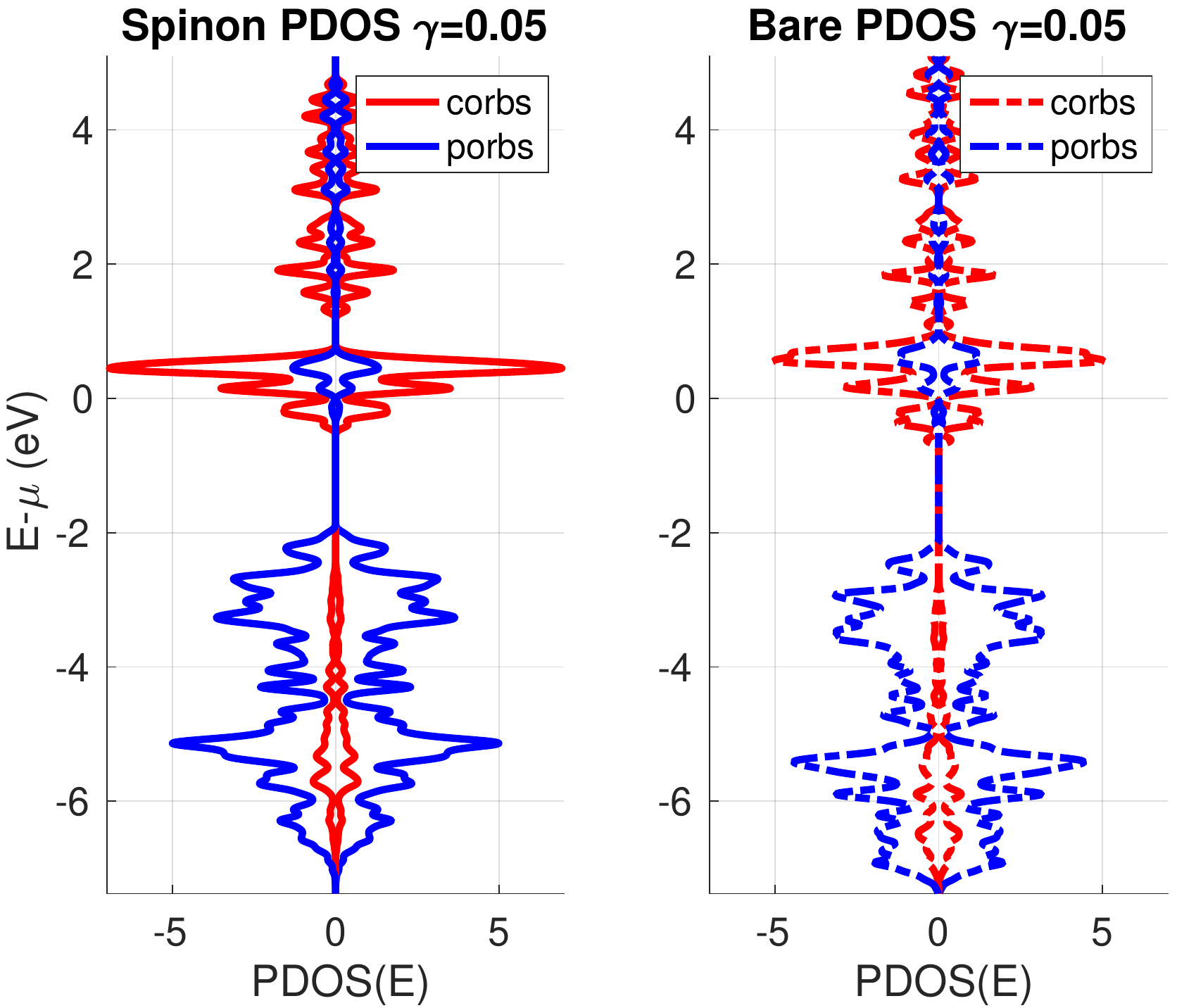}
\caption{Output of BoSS run for SrVO$_3$ (Example 1).  The slave model is the ``orbital slave''  for the V 3$d$ Wannier orbitals: 5 slaves per V atom, one slave per spatial orbital, no  spin resolution for the slaves so allowed slave occupancies are \{0,1,2\}; the interaction strengths of $U=U'=12$ eV, and $J=0$ eV are used for the V 3$d$ orbitals.   The left panel shows the band structure for both spin channels (which are identical due to lack of spin polarization): green shows the original Wannier (DFT) bands at $U=U'=J=0$ and red shows the renormalized spinon bands.  The right panel shows the projected spinon density of states (PDOS) onto the $d$ (V 3$d$) orbitals and $p$ (O 2$p$) orbitals for both the original and renormalized spinon bands (negative PDOS refers to spin down and positive to spin up).  The Gaussian broadening for the PDOS has a standard deviation of 0.05 eV.
}
\label{fig:svospinonbands}
\end{centering}
\end{figure}

\begin{figure}[t!]
\begin{centering}
\includegraphics[width=4.5in]{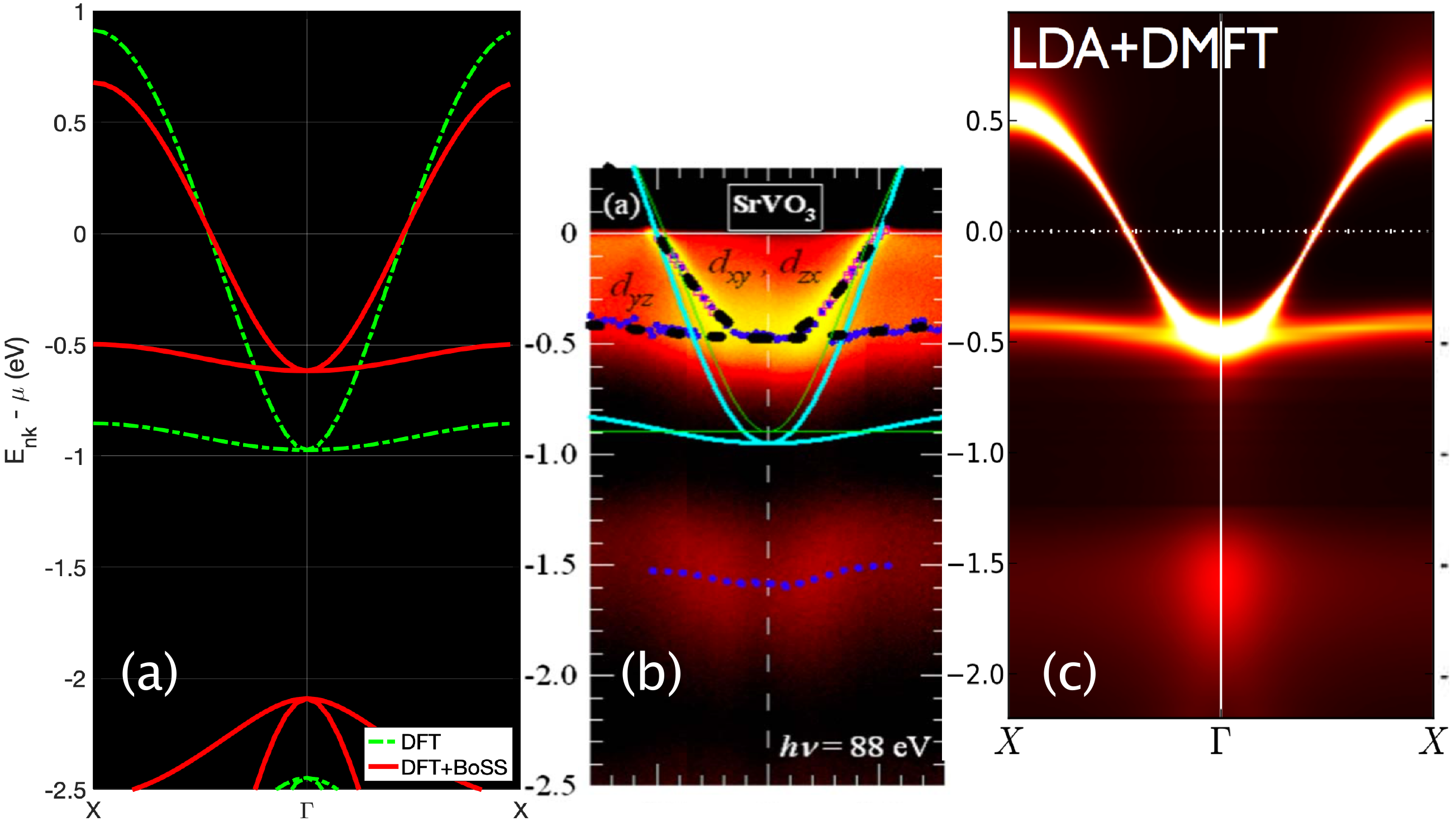}
\caption{Comparison of the electronic band structure of SrVO$_3$ from (a) BoSS, (b) experimental data from angular resolved photoemission spectroscopy (ARPES)~\cite{svoexpt}, (c) and DFT+DMFT~\cite{svodmft}.
The results in (a) show the non-interacting bands (dashed green labeled ``bare'') and the interacting spinon bands (solid red).  The BoSS calculations uses $U=12$ eV, $J=2$ eV, and $U'=8$ eV for the V $3d$ mainfold of orbitals, and ten slave modes are used per V atom (one slave per spatial orbital and spin combination) with allowed slave occupancies of \{0,1\}.
}
\label{fig:svobossexptdmft}
\end{centering}
\end{figure}

\underline{Example 1}:  Bulk SrVO$_3$ has a cubic perovskite structure with a five atom primitive unit cell with no observed spin polarization or other symmetry breaking.  A $p$-$d$ model is used with O 2$p$ and V 3$d$ Wannier orbitals (14 orbitals per unit cell). The full tutorial files include details of the DFT calculations including input files for the Quantum Espresso DFT package~\cite{QE} as well as the Wannier90 input file and output \verb+SVO_hr.dat+ tight-binding description.  Running the tutorial produces the band structure and projected densities of states (PDOS) shown in Figure~\ref{fig:svospinonbands}.  The main observation is that the spinon bands for the V 3$d$ conduction bands (those crossing the chemical potential $\mu$) become systematically narrowed in energy compared to the bare DFT bands, which corresponds to an effective mass enhancement by a factor of $\approx$ 2.  This is the primary effect of the local electronic interaction on the conducting electronic bands.  Figure~\ref{fig:svobossexptdmft} shows a direct comparison of BoSS electronic spectra to available experimental and DMFT data: as no effort at fine-tuning of the parameters was performed in the BoSS calculation, the comparison to prior work is very encouraging.

\underline{Example 2}:  The aim of this example is to show how easy it is to create a tight-binding representation file \verb+<base>_hr.dat+ by hand and thus create a manually specified Hubbard model.  The example creates a simple one-dimensional chain of alternating $d$ and $p$ sites.  Interested readers can examine the software package files for this example

\begin{figure}[t!]
\begin{centering}
\includegraphics[width=2.3in]{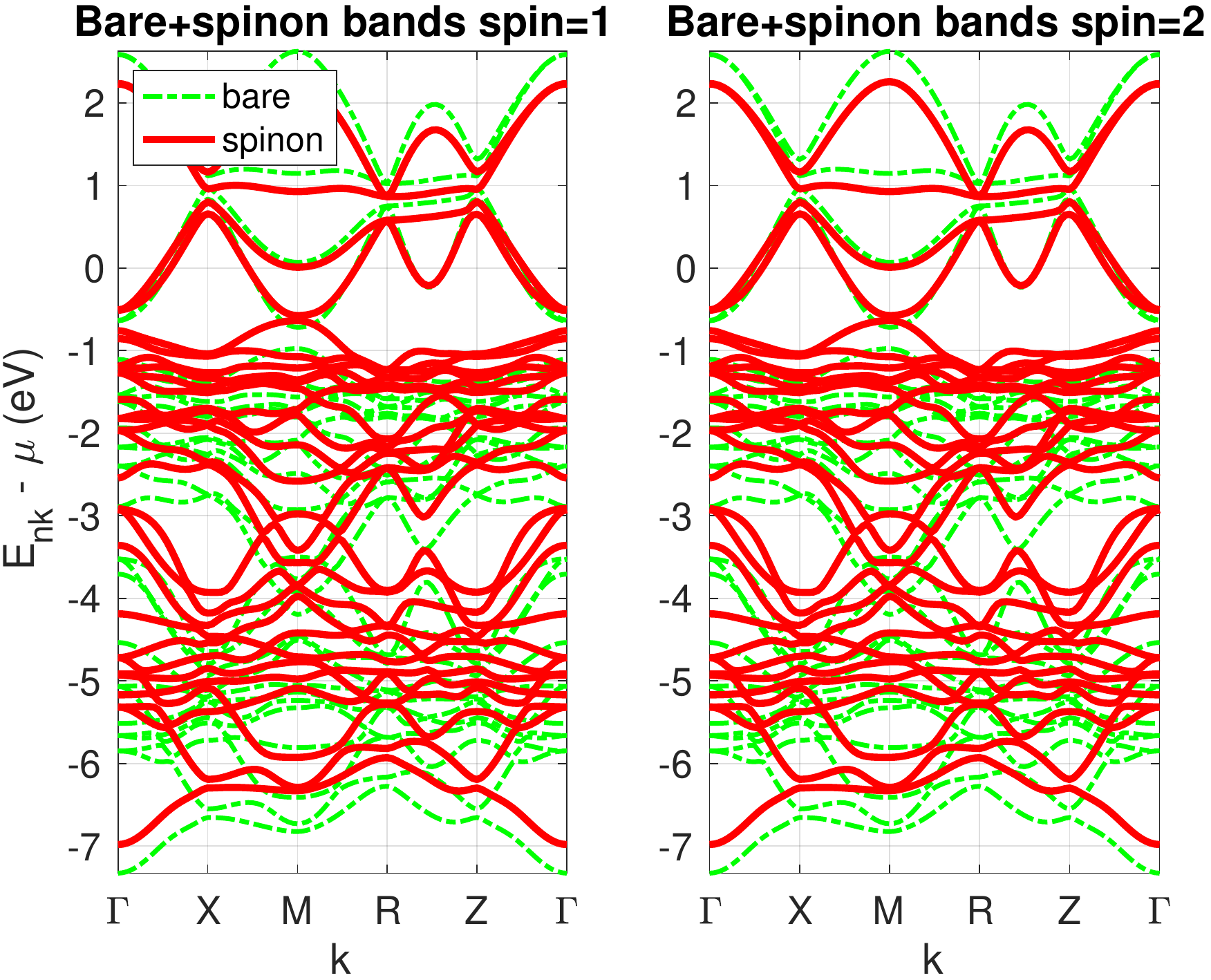}
\includegraphics[width=2.3in]{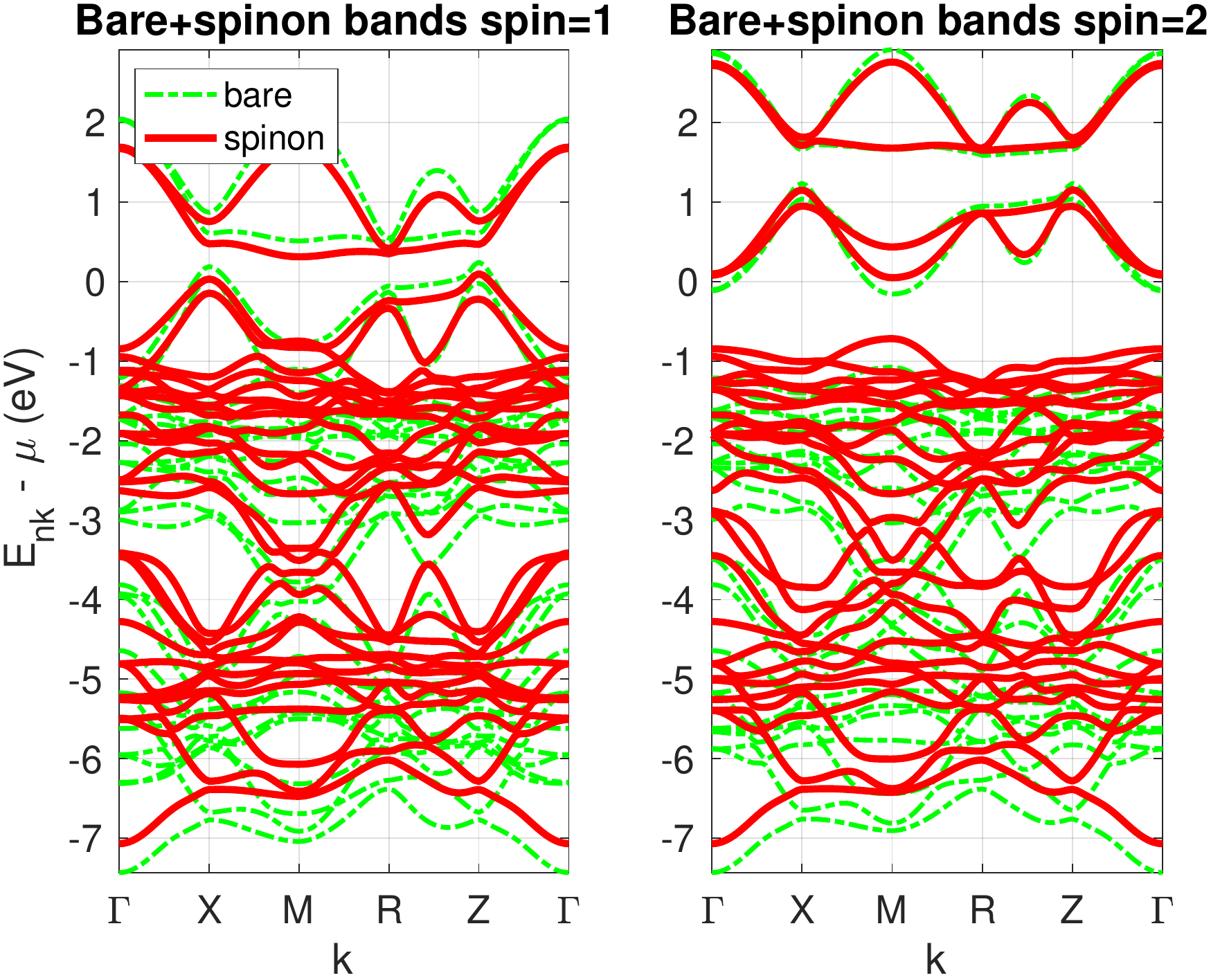}
\caption{Band structure output of BoSS run for SmNiO$_3$ (Example 3) run with $U=12$ eV, $U'=8$ eV and $J=2$ eV.  The simulation cell has two formula units (10-atom cell) with two inequivalent Ni sites.  Only the Ni $e_g$ orbitals (two per Ni) are treated as correlated (``$d$'') orbitals with the remaining $t_{2g}$ Ni 3$d$ orbitals and O 2$p$ orbitals treated as uncorrelated (``$p$'').  The left two panels show the electronic bands resulting without any magnetism ($B_{im\sigma}=0$): the two spin channels are necessarily identical and the system is metallic (incorrect compared to experiment).  The right two panels show the energy bands of the minimal $E_{tot}$ system with optimal $B_{im\sigma}$: the majority spin channel (spin=1) has two filled $e_g$ bands while all $e_g$ bands for minority spins (spin=2) are above $\mu$ and empty.  The spinon bands correctly predict an insulating material.}
\label{fig:snoB0Bmin}
\end{centering}
\end{figure}

\underline{Example 3}: SmNiO$_3$ is a perovskite-structured material with an insulating and antiferromagnetic ground state whose unit cell contains 80 atoms.  Instead of describing the full complexity of this system, this tutorial example focuses on a simpler description based on a 10-atom unit cell (two formula units) containing two inequivalent Ni cations: a ``breathing mode'' distortion exists in this material at low temperatures whereby one Ni atom has a larger oxygen octahedron surrounding it while the other Ni has a smaller octahedron.  This distortion is accompanied by a transition from a non-magnetic metal at high temperature to an insulating and magnetic system at low temperatures.  The magnetic struture in this small uit cell is taken as ferromagnetic for simplicity.  The tutorial files provides details of calculations with $B_{im\sigma}=0$, $B_{im\sigma}\ne0$ as well as the minimization over $B_{im\sigma}$ that yields the final optimal state of the system.  Here we will simply compare the magnetic and non-magnetic solutions.  Figure~\ref{fig:snoB0Bmin} compares the band structure of the two extremes: the optimized description with lowest $E_{tot}$ is insulating and magnetic, in agreement with experiment (the non-magnetic calculation is metallic).  Regardless of the magnetic state, the interactions reduce the width of the energy bands.

\underline{Example 4}: LaNiO$_3$ is a conducting transition metal oxide in which electronic interactions are known to lead to quantitative and observable changes of the electronic bands.  We choose a simple cubic unit cell for LaNiO$_3$ (one formula unit), which is the simplest representation and also allows for direct comparison to prior DMFT calculations.  
\begin{figure}[t!]
\begin{centering}
\includegraphics[width=3in]{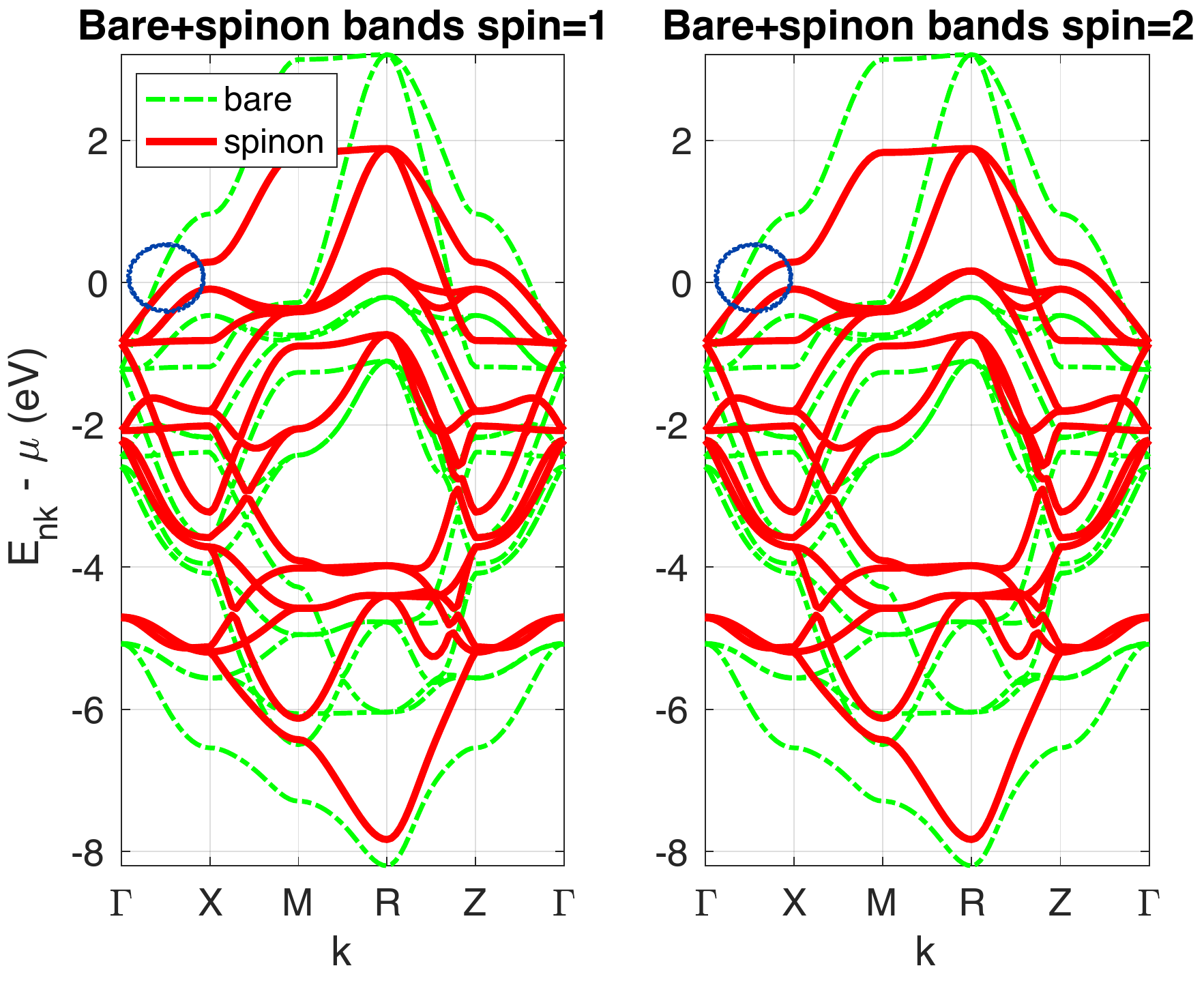}
\caption{Band structure output of BoSS  for cubic LaNiO$_3$ (Example 4) run with $U=10$ eV, $U'=6$ eV and $J=2$ eV without any magnetism.  There are two spatial correlated orbitals on the Ni (the $e_g$ orbitals which are $d3z^2-r^2$ and $dx^2-y^2$) and all remaining Ni and O 2$p$ orbitals are in the uncorrelated set.  Bare green dashed bands are the DFT-LDA results, and the solid red curves as the BoSS results for a slave model with full orbital and spin resolution (i.e., 4 slave modes, one for each spin and orbital combination with occupancies of either 0 or 1).  The two high energy bands of $e_g$ character cross the chemical potential $\mu$: the crossing in the $\Gamma-X$ direction is highlighted by the blue circle, and the slope of the bands at the crossings are the velocities $v_F^9$ (bare bands) and $v_F$ (spinon bands).}
\label{fig:lnobands}
\end{centering}
\end{figure}

The electronic bands for this system are displayed in Figure~\ref{fig:lnobands}.  We see that electronic interactions have a strong quantitative effect on the energy bands and make them narrower when compared to the non-interacting (bare) bands.  To quantify this effect, it is customary to compute ratios of the slopes of the bands (called the Fermi velocities, $v_F$) as they cross the chemical potential: the interaction reduces band width and thus the slope, and the ratio of the non-interacting to interacting slope, $v_F^0/v_F$, is often quoted as the ``effective mass enhancement factor'' and as a measure of the effect of electronic interactions and correlations on the energy bands.

\begin{table}[t!]
\begin{tabular}{cc|cccc}
 &  & \multicolumn{4}{c}{$J$ (eV)}\\
  &   $v_F^0/v_F$ & 0 & 1 & 2 & 3 \\
     \hline
          & 10 & 1.31 & 1.30 & 1.33 & 1.36 \\
$U$ (eV)  & 12 & 1.44 & 1.43 & 1.49 & 1.58 \\
          & 14 & 1.58 & 1.59 & 1.68 & 1.80 \\
\end{tabular}
\caption{Renormalization of the Fermi velocity along the $\Gamma-X$ direction for cubic LaNiO$_3$: $v_F^0$ is the DFT-LDA value and $v_F$ is the value for the spinon energy bands calculated by BoSS using the $U$ and $J$ values listed ($U'=U-2J$ throughout).  The slave model used has two slave modes on the Ni site representing the two $e_g$ Ni 3$d$ orbitals ($d3z^2-r^2$ and $dx^2-y^2$), and each slave mode can have occupancies in the set $\{0,1,2\}$ (no explicit resolution of the spin degree of freedom).}
\label{tab:vfnloorbslave}
\end{table}
\begin{table}
\begin{tabular}{cc|cccc}
 &  & \multicolumn{4}{c}{$J$ (eV)}\\
  &   $v_F^0/v_F$ & 0 & 1 & 2 & 3 \\
     \hline
          & 10 & 1.50 & 1.78 & 1.35 & 2.99 \\
$U$ (eV)  & 12 & 1.64 & 1.99 & 2.73 & 3.49 \\
          & 14 & 1.78 & 2.18 & 2.99 & 3.84 \\
\end{tabular}
\caption{Renormalization of the Fermi velocity along the $\Gamma-X$ direction for cubic LaNiO$_3$.  The nomeclature is identical to Table~\ref{tab:vfnloorbslave}, but the slave model used has four slave modes on the Ni site, one slave for each unique combination of spin channel and spatial $e_g$ orbital; each slave mode can  have occupancies in the set $\{0,1\}$.}
\label{tab:vfnlspinoorbslave}
\end{table}

\begin{figure}[t!]
\begin{centering}
\includegraphics[width=4.0in]{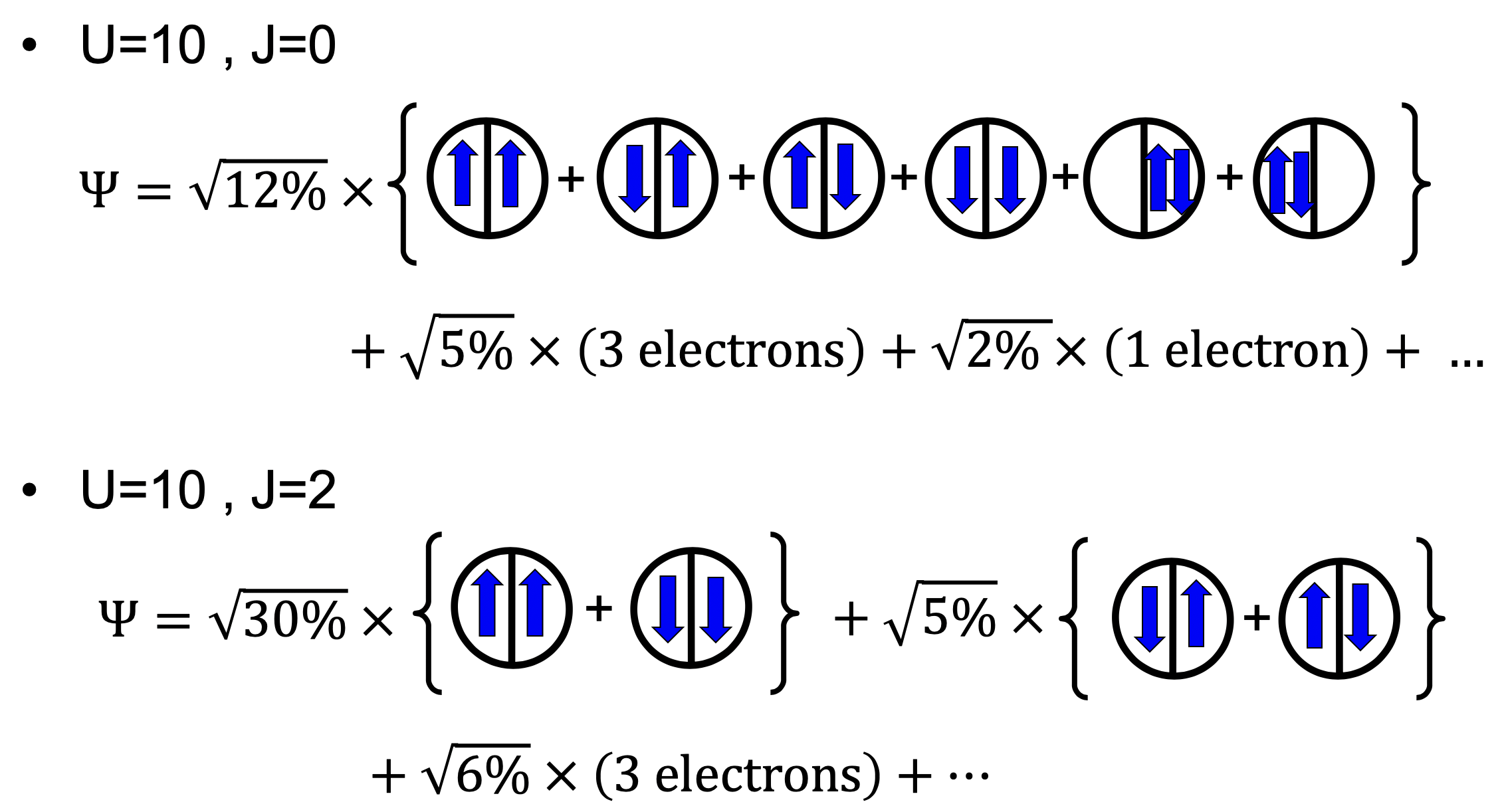}
\caption{Comparison of the spin+orbital slave ground state wave function of the Ni $e_g$ subsystem in cubic LaNiO$_3$.  The figure shows results for $U=U'=10,J=0$ (top part) and $U=10,U'=6,J=2$ (bottom part) where all parameters are in eV.  Each wave function is written as a superposition of configurations of the $e_g$ manifold: each circle represents a configuration; the left and right side of each circle represent the two $e_g$ spatial orbitals; the occupancy of each orbital is indicated by the presence (or absence) of blue arrows which also describe the spin occupancy.  The top wave function is dominated by two-electron configurations which then equally sample the orbitals and spin states (the remaining configurations have much smaller amplitudes).  The bottom wave function has a strong preference for spin-aligned two electron configurations due to the Hund's coupling.
}
\label{fig:nislavestates}
\end{centering}
\end{figure}

Tables~\ref{tab:vfnloorbslave} and \ref{tab:vfnlspinoorbslave} show the dependence of this slope ratio on the interaction parameters $(U,U',J)$ for two different slave models.  Experimental measurements~\cite{eguchi_fermi_2009} and prior theoretical work~\cite{deng_hallmark_2012} find that the ratio is approximately 3. As the tables show, this numerical value is achievable by fine-tuning the parameters in one of the slave models but not the other.  One of the features of BoSS is that it permits one to compare the two slave models in detail to see which physical effects create the the difference and lead to a better description of the actual material.  For example, looking at the two tables, why does the spin+orbital description generate larger, and more physically reasonable, mass renormalizations that are quite sensitive to the $J$ value?  To answer this, we can compare two spin+orbital calculations done with $\{U=U'=10,J=0\}$ and with $\{U=10,U'=6,J=2\}$ (all in eV).  Upon examining the interacting slave ground state for these two cases, we find the wave functions illustrated graphically in Figure~\ref{fig:nislavestates}.

As the figure shows, when $J=0$ and $U'=U$, the ground state has no preference between the different two-electron configurations: the system fluctuates between all six possible two-electron configurations equally and then rarely visits configurations with fewer or more electrons.  However, once $J>0$, this two-electron and two-orbital system can lower its energy by favoring the two spin-aligned configurations at the expense of other configurations: this greatly reduces the configurational fluctuations which in turn suppresses tunneling between Ni sites and thus the velocity of electron motion in the associated energy bands.  These effects have been described in prior literature as a feature of ``Hund's metals'' \cite{Yin2011,Hirjibehedin2015,PhysRevLett.117.247001,hundscouplingAGLdMMJ2013}.  What we are highlighting is the ease with which the BoSS approach allows one to identify the basic physics by suppressing or enhancing the mechanism via changes in the slave model: e.g., the results in  Table~\ref{tab:vfnloorbslave} are much less sensitive to the interaction parameters when compared to those in  Table~\ref{tab:vfnlspinoorbslave} because the former has no explicit description of the electron spin state and thus no way of selecting the spin-aligned configurations.

\section{Outlook}

The existing BoSS framework described in this paper is easy to modify and test.  Hence, it should be applied to a broad range of interacting electron systems to understand its performance, strengths, and limitations in terms of correctly predicting materials properties.  With the software available in open source form, accomplishing this important task is up to the theoretical materials physics community.

In terms of improved methodology and capabilities for the future, we identify a number of them in order of increasing difficulty.  First, the current software assumes that all the correlated atomic sites must have identical slave-boson models (i.e., the same slave $\alpha$ indices).  This limitation is easy to address by creation of improved data structures to handle each site separately.  Fortunately, the software already permits site-dependent values of the $U,U',J$ parameters.

Second, at present the software computes and reports the total energy $E_{tot}$, the mean occupations $\avg{\hat f^\dag_{im\sigma}\hat f_{im\sigma}}_f$ and $\avg{N_{i\alpha}}_s$ as well as the full spinon density matrix $\avg{\hat f^\dag_{im\sigma}\hat f_{i'm'\sigma}}_f$, and, spectroscopically, the spinon energy bands and projected densities of states.  Direct comparison to experimental spectroscopies, however, requires computation of the electron spectral function (of which the spinon energy bands form only one part).  Since both the spinon and slave eigenstates are computed by BoSS, all the required inputs to computing the spectral function within a slave-boson formalism are available in principle.  In practice, additional code and data structures must be implemented for the calculation of the spectral function after the BoSS solution is found.

Third, and more ambitiously, it is preferable to relax the current reliance on having the electron spin index $\sigma$ as an quantum number for electrons.  While this does permit the description of magnetic systems with collinear magnetic ordering, it does not permit arbitrary magnetic states or the description of spin-orbit coupled materials where the spatial ($m$) and spin ($\sigma$) degrees of freedom are necessarily mixed.  This will require reorganization of key data structures and more significant modification of the software stack.  A BoSS framework that can describe spin-orbit coupled electrons will enable a more realistic handling of materials containing 4$d$ and 5$d$ transition metal atoms.

\section*{Acknowledgement}

The initial development of BoSS was supported primarily by the National Science Foundation via the grant NSF MRSEC DMR-1119826. The Flatiron Institute is a division of the Simons Foundation. A. B. G. also acknowledges discussions with A. J. Millis and H. U. R. Strand.

\bibstyle{unsrt}
\bibliography{main}

\end{document}